\documentclass[symmetry,article,accept,moreauthors,pdftex]{Definitions/mdpi} 

\firstpage{1} 
\pubvolume{1}
\issuenum{1}
\articlenumber{0}
\pubyear{2021}
\copyrightyear{2020}
\datereceived{} 
\dateaccepted{} 
\datepublished{} 
\hreflink{https://dx.doi.org/} 

\usepackage[lined, linesnumbered, ruled]{algorithm2e}
\usepackage{xcolor}

\Title{Energy and Spectral Efficiency Balancing Algorithm for Energy Saving in LTE Downlinks}
\TitleCitation{\textls[-20]{Energy and Spectral Efficiency Balancing Algorithm for Energy Saving in LTE Downlinks}}

\Author{Mamman Maharazu $^{1,}$*\orcidA{}, Zurina Mohd Hanapi $^{2,}$*\orcidB{} and Mohamed A. Alrashah $^{2,}$*\orcidC{}}

\AuthorCitation{\textls[-20]{Maharazu, M.; Hanapi, Z.M.; Alrashah, M.A.}}

\AuthorNames{Mamman Maharazu, Zurina Mohd Hanapi and Mohamed A. Alrshah}

\address{%
	$^{1}$ \quad Department of Computer Science, Federal College of Education, College of Education, Katsina P.M.B. 2041 Katsina State, Nigeria
	$^{2}$ \quad Department of Communication Technology and Network, Universiti Putra Malaysia, Serdang 43400 UPM, Selangor D.E., Malaysia
}

\corres{Correspondence: maharazu2003@yahoo.com (M.M.); zurinamh@upm.edu.my (Z.M.H.); mohamed.asnd@gmail.com (M.A.A.)}

\abstract{In wireless network communication environments, Spectral Efficiency (SE) and Energy Efficiency (EE) are among the major indicators used for evaluating network performance. However, given the high demand for data rate services and the exponential growth of energy consumption, SE and EE continue to elicit increasing attention in academia and industries. Consequently, a study of the trade-off between these metrics is imperative. In contrast with existing works, this study proposes an efficient SE and EE trade-off algorithm for saving energy in downlink Long Term Evolution (LTE) networks to concurrently optimize SE and EE while considering battery life at the Base Station (BS). The scheme is formulated as a Multi-objective Optimization Problem (MOP) and its Pareto optimal solution is examined. In contrast with other algorithms that prolong battery life by considering the idle state of a BS, thereby increasing average delay and energy consumption, the proposed algorithm prolongs battery life by adjusting the initial and final states of a BS to minimize the average delay and the energy consumption. Similarly, the use of an omni-directional antenna to spread radio signals to the user equipment in all directions causes high interference and low spatial reuse. We propose using a directional antenna instead of an omni-directional antenna by transmitting signals in one direction which results in no or low interference and high spatial reuse. The proposed scheme has been extensively evaluated through simulation, where simulation results prove that the proposed scheme is efficiently able to decrease the average response delay, improve SE, and minimize energy consumption.}

\keyword{battery life; delay; energy-efficient; power saving; downlink; LTE}

\begin{document}

\section{Introduction}
Long Term Evolution (LTE) is an emergent wireless network technology that aims to provide low latency, peak data rate, wide coverage, and seamless mobility for different traffic types ranging from residential users to small businesses \cite{atayero20113gpp}. The mobility features that enable User's Equipment (UE) to move at vehicular speeds are included in the first LTE (Release 8), followed by several incremental improvements in Releases 9 and 10~\cite{holma2012lte}. The mobility of UEs poses numerous issues because the UE is not only powered by a rechargeable battery, but also has a narrow capacity for providing the required consumption. Consequently, an efficient power-saving scheme is necessary to prolong the battery lifetime of the UE at a radio Base Station (BS), denoted as eNodeB, before recharging. 

The discontinuous reception (DRX) power-saving algorithm \cite{tung2015analysis} adjusts the idle threshold to improve the Energy Efficiency (EE) and Spectral Efficiency (SE) of eNodeB. Such adjustment is overwhelmed by the remaining energy at eNodeB. Consequently, eNodeB quickly transits into sleep-mode to save energy, which increases average response delay because eNodeB has to wait for the next listening window to receive a message. The idle threshold is adjusted using exponential and general distribution methods, which yields frequent transitions into the listening mode when the packet arrival rate at eNodeB is small, where frequent transitions may lead to high energy consumption. Moreover, dynamically adjusting the idle threshold is difficult because of time-varying traffic. Therefore, an efficient SE and EE Trade-off (SET) algorithm is proposed in this study; this algorithm mitigates the ineffectiveness of the DRX algorithm. The major difference between the proposed SET algorithm and the DRX algorithm is the manner in which the sleep-mode is adjusted: initial and final state. The packet arrival interval and transmission time of the DRX algorithm are adjusted using exponential and general distribution methods, respectively, whereas that of the proposed SET algorithm are adjusted in accordance with the packet arrival pattern.

In this study, an enhancement to the DRX power-saving algorithm, namely SET, is proposed for saving energy in downlink LTE networks. The Radio Resource Control (RRC) idle state in the DRX algorithm is analytically enhanced. In contrast with the DRX algorithm, the proposed SET algorithm dynamically adjusts the RRC at initial and final states on the basis of downlink stochastic arrival pattern. In addition, an improved sleep-mode scheme is proposed to minimize numerous transitions to listening mode, which consequently reduces the high energy consumption when the traffic arrival is less. The simulation results show that the SET algorithm considerably outperforms the DRX algorithm in terms of SE, SET, average response delay, and energy consumption.

The major contributions of this study are fourfold. Firstly, SE and EE are concurrently optimized as a Multi-objective Optimization Problem (MOP). Secondly, the MOP is transformed to a Single-objective Optimization Problem (SOP) by using a weighted sum technique, thereby proving that the scheme is quasi-concave while its Pareto optimal solution is also examined. Thirdly, an efficient algorithm namely SET is proposed to prolong the battery life, which reduces energy consumption at a BS. Lastly, the performance of the proposed SET algorithm is evaluated via simulation, where the results clearly demonstrate the considerable gain of the SET algorithm compared to other methods.

The rest of this paper is organized as follows: Section \ref{LR} provides an overview of related works, Section \ref{SET} presents the proposed SET algorithm, Section \ref{PE} illustrates the performance evaluation and the benchmark of the proposed SET algorithm, and finally, Section \ref{Con} concludes the entire work.

\section{Related Work} \label{LR}
Several algorithms proposed for LTE downlinks are reviewed in this section. The review focuses on how these algorithms use SE, EE, SET and sleep-mode operations to save energy, where the advantages and limitations of each algorithm are highlighted. Recently, several studies have been conducted to investigate the relationship between SE and EE. In \cite{tang2014resource}, a new paradigm was presented for SET by considering different bandwidth requirements. The scheme analyzed resource efficiency for Orthogonal Frequency-Division Multiple Access (OFDMA) and proved that it utilizes the trade-off between EE and SE by balancing the power consumption and the occupied bandwidth. However, this scheme did not provide an appropriate weight factor to utilize the available bandwidth and power. Later, the trade-off between EE and SE in a delay-constrained wireless system was proposed in \cite{chen2015tradeoff}. The approach used a generic closed-form approximation to investigate the impact of Quality of Service (QoS) and circuit power consumption on EE and SE, respectively, by applying a curve-fitting mechanism. This approach proved that EE function is a quasi-convex problem that can be solved using a two-step binary search algorithm. However, the scheme lacked a closed-form power allocation approach and a mathematical formulation for the trade-off between EE and Effective Capacity (EC). SET trade-off behavior has been widely considered using various interference-level scenarios. 

SET with filter optimization in multiple access was analyzed in \cite{souza2015energy} using two conflicting metrics: throughput maximization and power consumption minimization. An energy-efficient design for an OFDMA network was proposed in \cite{abrao2016energy} based on Wu’s effective capacity approach to maximize system throughput, which is subject to delay QoS requirements. Furthermore, effective EE and EC trade-off were utilized through fractional programming, which converted quasi-concave optimization to a subtractive optimization problem by adopting Dinkelbach’s technique. However, this algorithm was not optimal for maximizing EC-based statistical delay provisioning.

In \cite{son2015energy}, a general problem was formulated to minimize total consumption cost while satisfying area SE requirements. The problem was further decomposed into a deployment problem at peak time and an operational problem at off-peak time. However, network performance was jeopardized due to the high energy consumption of the scheme. In \cite{song2015energy}, a MOP was adopted to examine SET in downlink OFDMA systems with fairness constraints. The scheme uses a weighted sum method to obtain Pareto sets that provides a quantitative insight into SET with various fairness levels. However, the scheme did not achieve trade-off between EE and fairness because the backhaul energy consumption was completely disregarded.

Reference \cite{pervaiz2015energy} proposed a scheme that jointly maximizes overall system EE and SE towards green heterogeneous networks under QoS constraints. The MOP was modeled to dynamically adjust the trade-off parameters of network providers. To obtain a Pareto optimal solution, the problem was transformed into an SOP using a weighted sum method. However, this scheme disregarded the total signaling power of the transmitters, transmission time and power constraints of the antennas. In \cite{li2015energy}, SET was investigated for a downlink OFDMA single-cell network. The scheme simultaneously optimized EE and SE, which were then formulated as a MOP. It used a weighted linear sum technique to convert the MOP to an SOP where Pareto optimal sets were analyzed. The converted SOP illustrated that the scheme is neither quasi-concave nor quasi-convex. Thereafter, the scheme was solved using particle swarm optimization, which decreased the total transmission power and improved the EE. However, this scheme is neither convex nor concave, {thus, finding its optimal solution is difficult and also requires extra processing that consumes more power.}

In \cite{kim2008remaining}, the authors proposed an energy-aware power management scheme that dynamically updates sleep parameters in accordance with the remaining energy of Inter-Arrival Time (IAT) to prolong battery life. The scheme achieved minimum response delay under sufficient energy but resulted in high energy consumption under insufficient energy. Reference \cite{richter2009energy} investigated the EE aspects of the BS deployment mechanism for cellular networks. The mechanism explored the impact of power consumption on the basis of deployment strategies. In addition, the concept of area power consumption was introduced. The scheme achieved an average throughput improvement but resulted in high energy consumption. After that, a scheme for power allocation and antenna port selection in Orthogonal Frequency-Division Multiplexing (OFDM) distributed antenna systems was proposed in \cite{ling2010schemes}. Two schemes were investigated to maximize the downlink received signal. Firstly, the scheme selected the distributed antenna on the basis only of large-scale fading and power allocation. Secondly, the scheme selected the distributed antenna using large- and small-scale fading coupled with optimal power. The scheme achieved low complexity but disregarded delay performance, which resulted in a waste of energy.

Energy-efficient link adaptation with transmitter Channel State Information (CSI), which was proposed in \cite{isheden2011energy}, aimed to minimize the total energy consumption of a mobile terminal. Flat and frequency-selective fading channels were used to determine the consumed energy. The algorithm achieved a decrease in energy consumption with an increase in bandwidth, however, it could not guarantee the user’s QoS. In \cite{haider2012energy}, an energy-efficient subcarrier and bit allocation for multi-user OFDMA systems were proposed. SET was analyzed by considering fairness constraints among users. The algorithm was formulated as an optimization problem for integer fractional programming. Moreover, an iterative programming method was used to solve the optimization problem by minimizing system complexity. However, the algorithm resulted in poor energy efficiency because it assumed that CSI is always perfect. In fact, CSI can be corrupted when the scheduled data rate surpasses the maximum channel size.

The authors of \cite{khakurel2013trade} investigated SET in a fading communication link. An optimization problem to maximize ergodic SE with a constraint of minimum ergodic EE was introduced. The scheme achieved an improvement in SE while minimizing channel circuit power, but it did not consider the impact of the user’s data rate, which resulted in inefficient resource utilization. To alleviate this problem, Reference \cite{coskun2017energy} proposed SET for a heterogeneous network with QoS constraints. This scheme aimed to simultaneously maximize SE and EE while satisfying the data rate requirement of each user. It was designed on the basis of three stages. Firstly, the cell center radius was selected using fractional frequency reuse. Secondly, frequency resources were allocated to satisfy user data requirements. Lastly, the Levenberg--Marquardt approach was used to solve the power allocation sub-problem. This scheme achieved higher outage probabilities because of increased inter-cell interference, however, it resulted in high energy consumption.

The authors of \cite{tsilimantos2016spectral} proposed SET in cellular networks. The algorithm developed a theoretical framework that is applicable to OFDMA networks based on transmission power and optimal resource allocation. Initially, the algorithm focused on using a single-cell scenario but was later extended to a multi-cell scenario using the stochastic geometry approach. It achieved tractable outcomes but using a large number of antennas resulted in high energy consumption. SET in an interference-limited algorithm was proposed in \cite{li2017energy}, which concurrently optimizes EE and SE. Firstly, the scheme was formulated as a MOP by applying the constraint of the transmission power limit. A weighted linear sum technique was used to convert the MOP to an SOP. The scheme achieved a balance between EE and SE but it led to inefficient use of network resources. 

The joint evaluation of the EE downlink scheduling algorithm was recently proposed in \cite{salman2018joint}. The algorithm aimed to optimize energy and bandwidth resources while guaranteeing QoS at the downlink by considering partial feedback. At eNodeB, SE and EE were optimized by amending the downlink scheduler in accordance with the packet prediction mechanism. The algorithm achieved an EE improvement of up to 79\% but generated considerable overhead, which resulted in high energy consumption.

In this paper, we propose a new energy-efficient algorithm, namely SET, to mitigate the aforementioned problems. The algorithm analytically adjusts the initial and final sleep windows by considering the downlink stochastic packet arrival pattern and introducing an improved sleep-mode. 

\section{The Proposed SET Algorithm} \label{SET}
This section introduces a new energy-efficient algorithm that prolongs battery lifetime at eNodeB, which is able to adaptively adjust the network parameters. 

\subsection{Energy-Efficient Battery Lifetime Algorithm}
Reference \cite{tung2015analysis} analytically investigated the efficiency of the DRX mechanism and the effect of RRC state transmission on saving battery life. The performance of this scheme has been evaluated by considering only idle-state RRC. However, this scenario causes an increase in packet delay that might result in high energy consumption.

The major difference between the proposed scheme and the DRX power-saving algorithm is the manner in which the sleep-mode of RRC is adjusted: the initial state $(initial_{state})$ and final state $(final_{state})$. Furthermore, the performance of the proposed algorithm is evaluated analytically and through experimental simulation. Moreover, packet arrival interval and transmission time in DRX are adjusted on the basis of exponential and general distributions, respectively. In contrast with DRX, the proposed SET algorithm dynamically adjusts three parameters on the basis of the packet arrival pattern. The basic concept behind the sleep operation is to improve power consumption by minimizing energy. In general, eNodeB goes into sleep-mode when UE has no request for packet processing. 

\begin{figure}[H]
	\centering
	\includegraphics[width=0.7\linewidth]{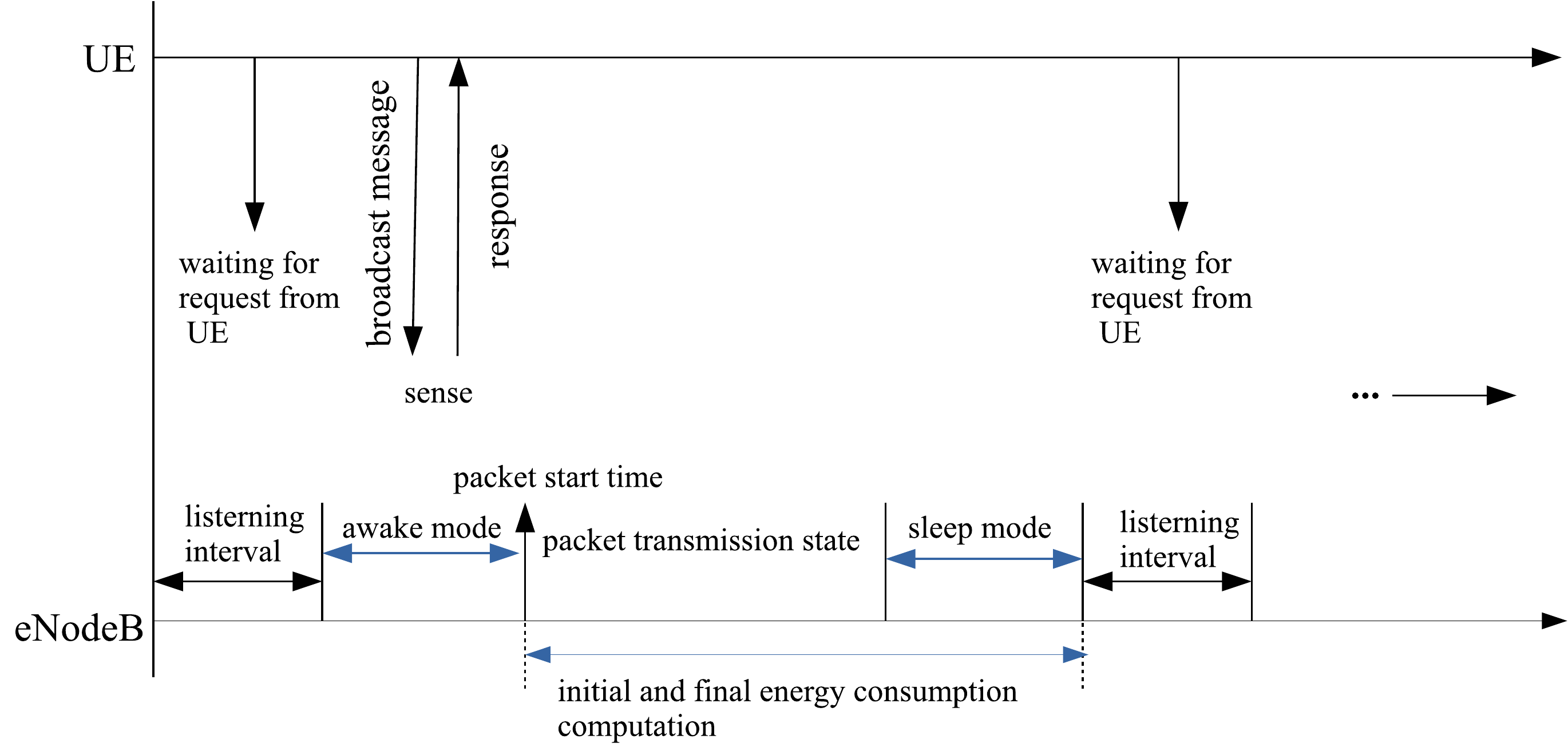}
	\caption{SET 
		algorithm operation with awake and sleep-modes.}
	\label{fig:sleepmodeSET}
\end{figure}

{The communication between UE and eNodeB with their corresponding sleep operation has been illustrated in Figure~\ref{fig:sleepmodeSET}}, where the eNodeB remains in the idle state unless it receives a message from UE that a packet has to be transmitted. Consequently, the eNodeB transits from a listening state to an aware mode by sensing and then responding to the UE that is ready to receive an incoming packet for transmission.

The time that the incoming packet arrived at the eNodeB scheduler and the time spent before it is being transmitted are recorded. When the current time ({spent-time}--{arrival-time}) is longer than 1 ms, the packet is in the transmission state. During the transmission state, the Initial Energy Consumption (IEC) and the Final Energy Consumption (FEC) are computed using Equations \eqref{eq1} and \eqref{eq2}, respectively. After all the packets are transmitted, eNodeB returns to sleep-mode while waiting for a request from the UE to transmit another incoming packet. This process is repeated continuously until an optimal value is obtained for IEC and FEC.  

Given that the remaining energy at eNodeB is consumed, the act of eventually saving energy is crucial. IEC has an impact on energy consumption and packet response delay, which should be adjusted by considering the remaining energy in eNodeB as indicated in Equation \eqref{eq1}:
\begin{equation} \label{eq1}
	IEC= max(\frac{E_{total}-E_{remained}}{E_{total}}) * E_{max}
\end{equation}
where $E_{total}$ denotes the total energy of eNodeB, $E_{remained}$ represents the remaining energy, and $E_{max}$ is the maximum energy. Similarly, FEC is calculated using Equation \eqref{eq2}:
\begin{equation} \label{eq2}
	FEC = a \times E_{max} + b \times E_{min}
\end{equation}
where $E_{min}$ is the minimum energy, and $a + b = 1$, which represents the weight of each value and bases the estimates of the current IAT on the past IAT.

\begin{figure} [H]
	\centering
	\includegraphics[width=0.6\linewidth]{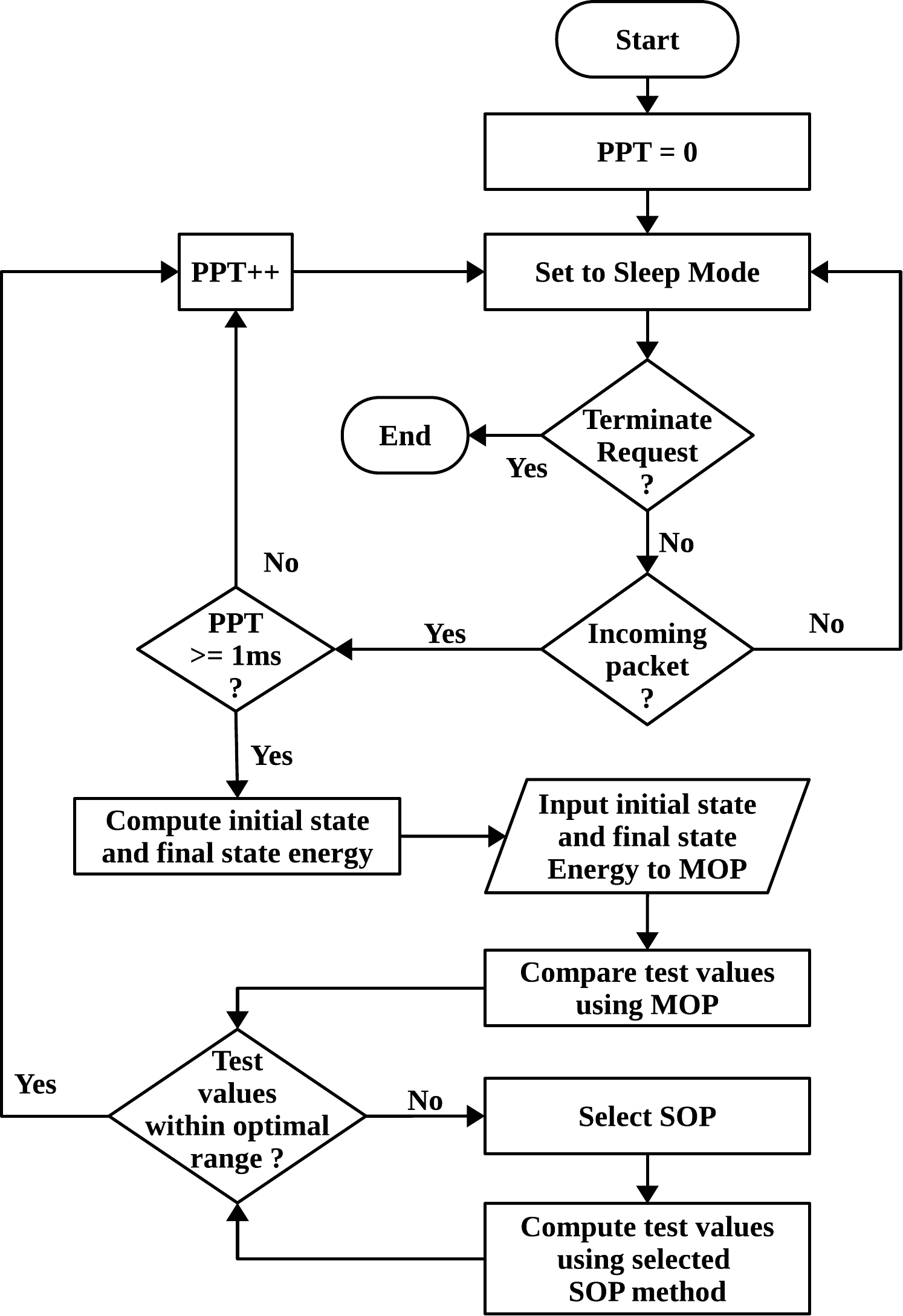}
	\caption{Control flow diagram of the SET algorithm. }
	\label{fig:SetFlowchart}
\end{figure}

{The primary modification and improvement of the proposed SET algorithm have been illustrated in Figure~\ref{fig:SetFlowchart}, in which the packet processing time is set to $\geq 1$ ms as described in 3GPP Release 9 \cite{LTE2010specification}}, and the initial and final energy states are introduced. 

The corresponding algorithm is presented in Algorithm \ref{algo01}, which illustrates how the two parameters are dynamically adjusted based on the downlink stochastic packet arrival pattern. Firstly, the idle, initial, and final states initialized at the UE, which begins in normal sleep-mode operation and waits for a short period; if no packet is received, then it sends a request message to the eNodeB to enter into sleep-mode.

The eNodeB computes the initial and final sleep windows for the UE using Equations \eqref{eq1} and \eqref{eq2}, and then sends the results back to the UE for it to proceed to sleep-mode. Thereafter, the request to terminate sleep-mode (i.e., request to wake up from sleep) is measured; if it is not satisfied, then the UE returns to normal operation mode; otherwise, the operation is terminated. 

\subsection{Determination of Pareto Optimal Solution}
The primary objective of this section is to concurrently optimize EE and SE to obtain an optimal solution. Assume that adaptive coding and modulation scheme is used to attain the Shannon rate limit, the SE of the $k^th$ UE can be illustrated as: 
\begin{equation} \label{eq3.1}
	SE(p_k) = \log_2(1 + \beta_k)	
\end{equation}
where $p_k$ denotes the transmission power and $\beta_k$ is the instantaneous signal-to-noise ratio of the $k^{th}$ UE, and $k \in \{1, \cdots, K\}$. While, the SE of all UEs can be represented as:
\begin{equation} \label{eq3.2}
	SE(P) = \sum\limits_{k=1}^{K}\log_2(1 + \beta_k)	
\end{equation}
where $P = \{p_1, p_2, \cdots, p_K\}$. Furthermore, EE is defined as the ratio of SE over the total power consumption, which can be described as:
\begin{equation} \label{eq3.3}
	EE = \frac{SE(P)}{\sum\limits_{k=1}^{K}(p_k + p'_k)}	
\end{equation}
where $p'_k$ represents the remaining percentage of circuit power of the $k^{th}$ UE.

Hence, the problem is formulated as a MOP, which can be described as in Equation~\eqref{eq3}:
\begin{equation} \label{eq3}
	MOP = max\left\{EE, SE\right\}	
\end{equation}
subject to $0 \le p_{k} \le p_{max}, \forall k \in K$, where $p_{max}$ denotes the maximum transmission power allocated to each eNodeB and $SE = SE(P)$ as in Equation \eqref{eq3.2}. 

To solve the SET in the aforementioned MOP, the concept of Pareto optimality \cite{emmerich2006multicriteria, coello2007evolutionary} is used, which results in the conversion of the MOP to an SOP using weighted linear summation technique \cite{marler2004survey}, as shown in Equation \eqref{eq5}:
\begin{equation} \label{eq5}
	SOP = (\theta \times SE) + ((1 - \theta) \times EE)
\end{equation}
subject to $0 \le p_{k}\le p_{max}, \forall k \in K$, where Equation \eqref{eq5} represents the SET optimization problem (objective function), and $\theta$ is the trade-off weighting parameter, such that \mbox{0 $\le$ $\theta$ $\le$ 1}, which enables the flexibility to achieve the trade-off between EE and SE.

\begin{algorithm}[h!]
	\caption{SE and EE Trade-off (SET) Algorithm for Saving Energy in Downlink LTE Networks}\label{algo01}
		
	\textbf{Input:}\\
		\hspace{1cm}$PPT$ : Packet processing time\\
		\hspace{1cm}$State$  : System State\\
		
	\textbf{Initialization:}\\
		\hspace{1cm}$PPT = 0$;\\
		\hspace{1cm}$State =$ {Sleep-mode};\\
		
	\While{State = Sleep-mode}
	{
		\uIf{termination request is received}
		{
			$State = Terminate$;
		}
		\Else
		{
			\If{incoming packet is available}
			{			
				\If{PPT $\geq$ 1}
				{
					$SE(p_k) = \log_2(1 + \beta_k)$;\\
					$SE(P) = \sum\limits_{k=1}^{K}\log_2(1 + \beta_k)$;\\
					$EE = \dfrac{SE(P)}{\sum\limits_{k=1}^{K}(p_k + p'_k)}$;\\
					$MOP = max\left\{EE, SE\right\}$;\\	
					$SOP = (\theta \times SE) + ((1 - \theta) \times EE)$;\\

					\While{SOP is not within optimal range}
					{
						Select $SOP$;\\
						Compute test values using $SOP$ method;\\
					}
				}
				{PPT++};
			}
		}
	}	
\end{algorithm}

\subsection{Energy Consumption of eNodeB Using Congested and Non-congested Network Loads}
\begin{figure}[H]
	\centering
	\includegraphics[width=0.5\linewidth]{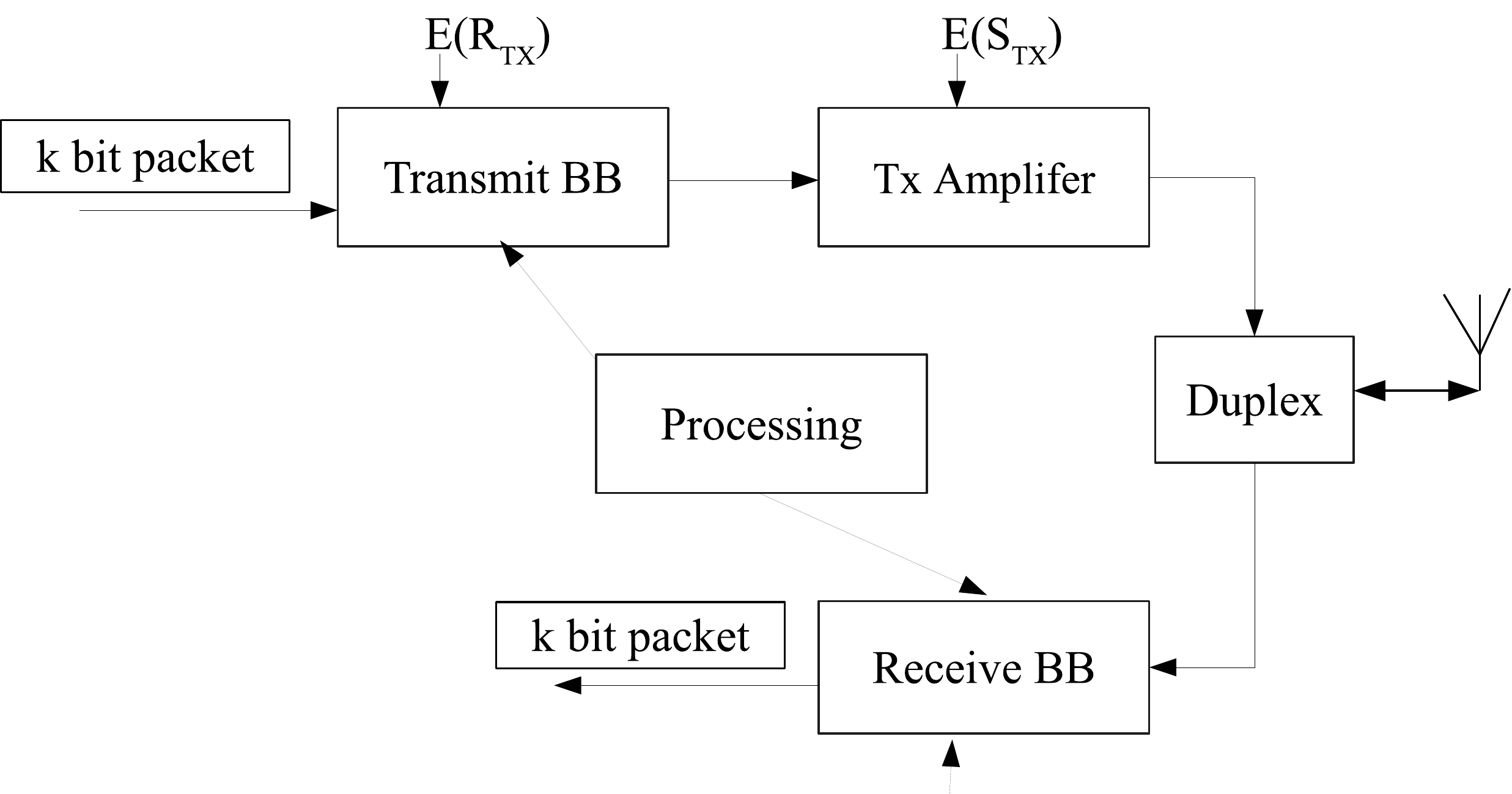}
	\caption{The adopted energy consumption model.}
	\label{fig:EnergyModel}
\end{figure}

Recently, global warming and $CO_{2}$ emission have been extensively studied from the perspectives of environmental and economic impacts \cite{lee2017distributed}. In this regard, energy consumption at eNodeB has become a key issue in cellular networks. Studies have determined that eNodeB operations are accountable for a major part of cellular energy consumption \cite{liu2018energy} wherein up to 80\% of energy is consumed. Therefore, minimizing the power consumption of eNodeB can considerably reduce overall network energy consumption. The scheme proposed in this section relies on the path loss and energy consumption of eNodeB. Accordingly, we adopt and modify the energy consumption model in references \cite{auer2011much, jensen2012lte}, as shown in Figure~\ref{fig:EnergyModel}.

From Figure~\ref{fig:EnergyModel}, the total energy consumption can be formulated as:
\begin{equation}
	E_{Total} = i(E(R_{TX}) + E(S_{TX}))
\end{equation}
where i denotes the number of users, $E_{Total}$ denotes the total energy consumption, $R_{TX}$ and $S_{TX}$ denote the transmission and receiving power levels measured in {i}Joules/bits, respectively.

\section{Performance Evaluation of the Proposed SET algorithm}\label{PE}
This section primarily aims to perform elaborate evaluations of the proposed SET algorithm to validate its performance compared to the DRX power-saving algorithm.

\subsection{Simulation Setup}
The simulation scenario in this section is based on two types of networks served by a single eNodeB; (1) A non-congested network with a number of UE less than 100, where the network density is under normal non-congested condition \cite{LTE2010specification}. (2) A congested network with a number of UE greater than 100 \cite{wang2016service}. The total energy consumed by the entire network is the summation of energy dissipated during the transmission and receiving processes, which comprises energy spent in the control and data messages in both modes.

\begin{figure}[H]
	\centering
	\includegraphics[width=0.4\linewidth]{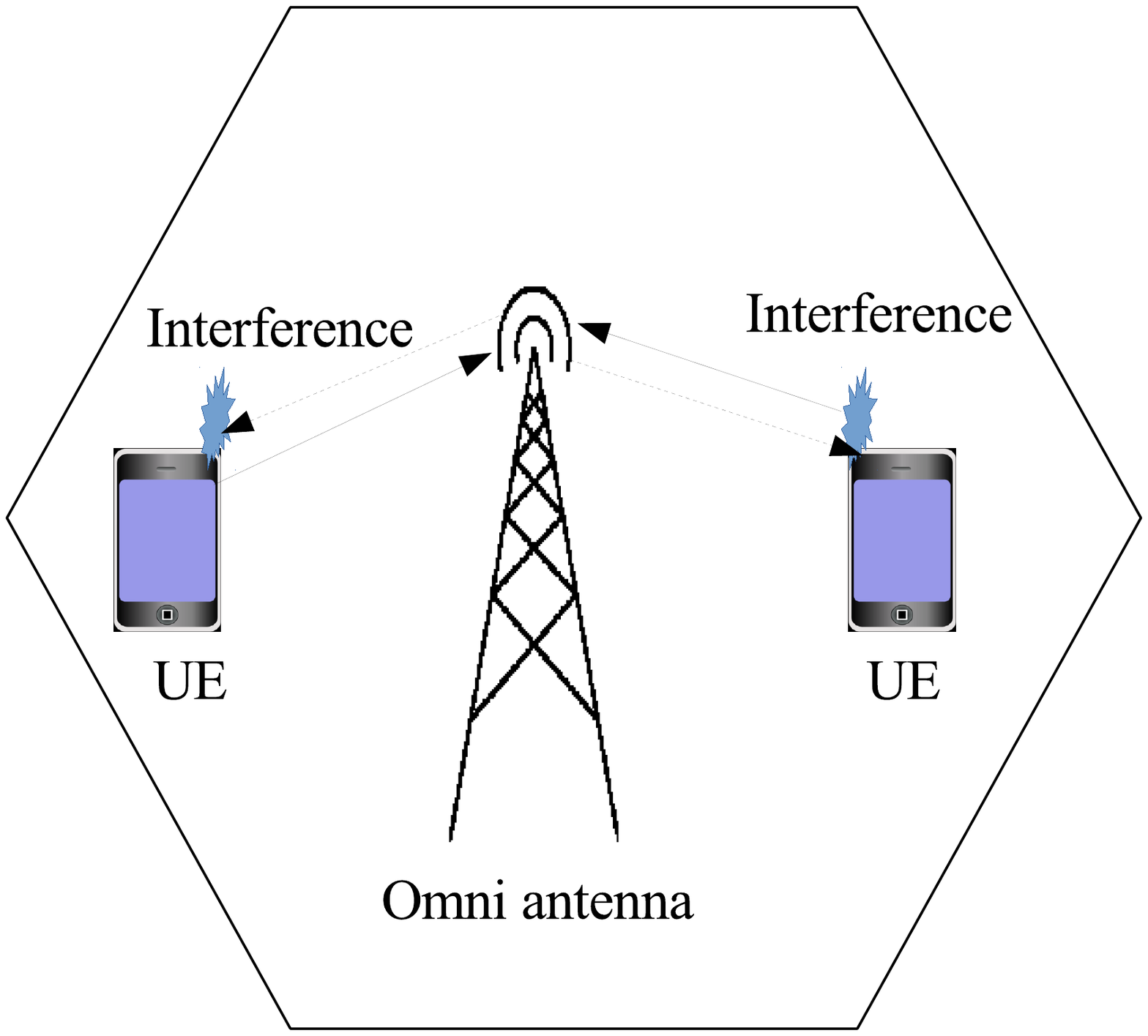}
	\caption{Omni-antenna in cellular network with interference.}
	\label{fig:Interference}
\end{figure}

A cellular network that comprises homogeneous macro-cells with a hexagonal tessellation is presented as in Figure~\ref{fig:Interference}. The eNodeB is placed at the center of each cell, and UE is assumed to be served by the nearby eNodeB. Spectrum resources and power are assumed to be allocated equally to all cells in the network. Two distinct eNodeB antenna configurations are used in this section: (1) omni-directional and (2) directional antennas. An omni-directional antenna transmits radio links in all directions but has limited network capacity because of the high interference. 

\begin{figure}[H]
	\centering
	\includegraphics[width=0.5\linewidth]{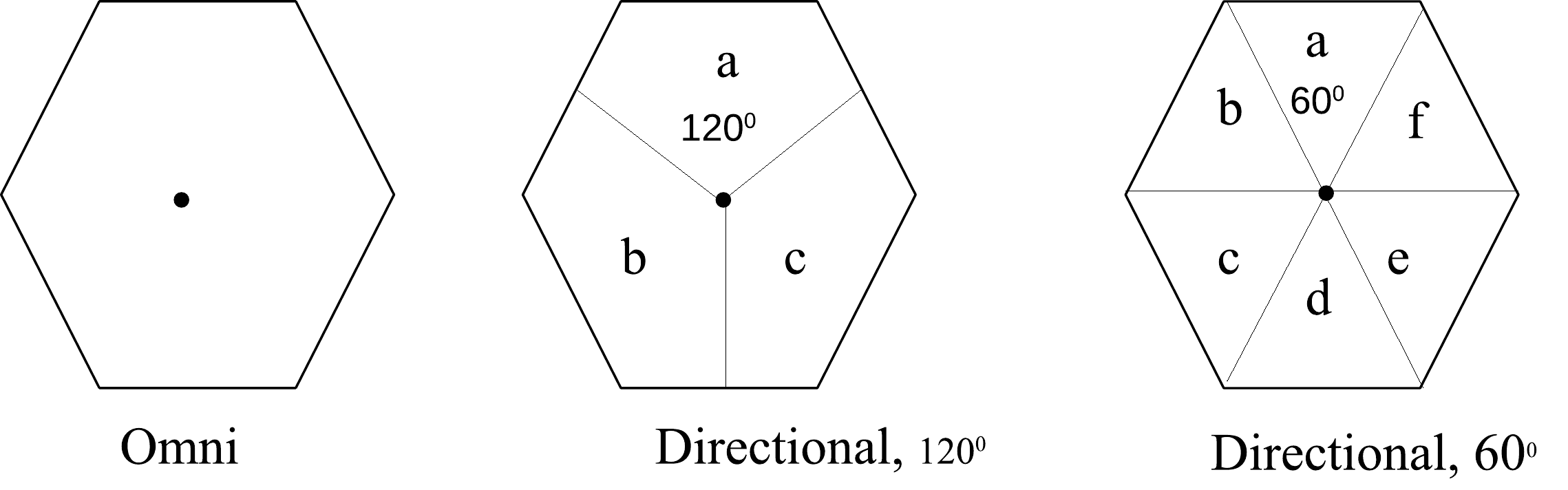}
	\caption{Antenna patterns.}
	\label{fig:Antenna_Pattern}
\end{figure}

To alleviate interference caused by an omni-directional antenna, cell sectoring, which divides cells into sectors (as $120^\circ$ or $60^\circ$) using a directional antenna, is proposed. The transmission and receipt of radio signals in a certain direction result in low interference, high spatial reuse, long transmission range and improved network capacity. The cell sectoring by antenna patterns has been shown in Figure~\ref{fig:Antenna_Pattern}. 

To evaluate the performance of any cellular network, an important metric that should be considered is the signal-to-interference-plus-noise ratio (SINR). Thus, the SINR of user's cells in a region served by cell $k$ is calculated using Equation \eqref{eg:sinr} based on Reference \cite{elwekeil2017performance}:
\begin{equation} \label{eg:sinr}
	SINR = \left( \frac{P_{k}*C_{k,n}}{\sum \limits_{i = 1, \text{ except } i = k}^{I} 		P_{i}*C_{i,n} + P_{n} }\right)
\end{equation}
where $P_{k}$ denotes the transmission power of the serving cell {k}, $C_{k,n}$ represent the channel gain from the serving cell $k$ to $n$ number of UE, $I$ is the set of all interfering cells in the region. Moreover, $P_{i}$ is the transmission power of each $i^{th}$ neighboring cell, $C_{i,n}$ represents the channel gain from the $i^{th}$ cell to $n$ number of UE, $P_{n}$ is the white Gaussian noise power, where the channel gain in this research considers both transmission and receiving antenna gains along with their path loss gain.

\begin{figure} [H]
	\centering
	\includegraphics[width=0.5\linewidth]{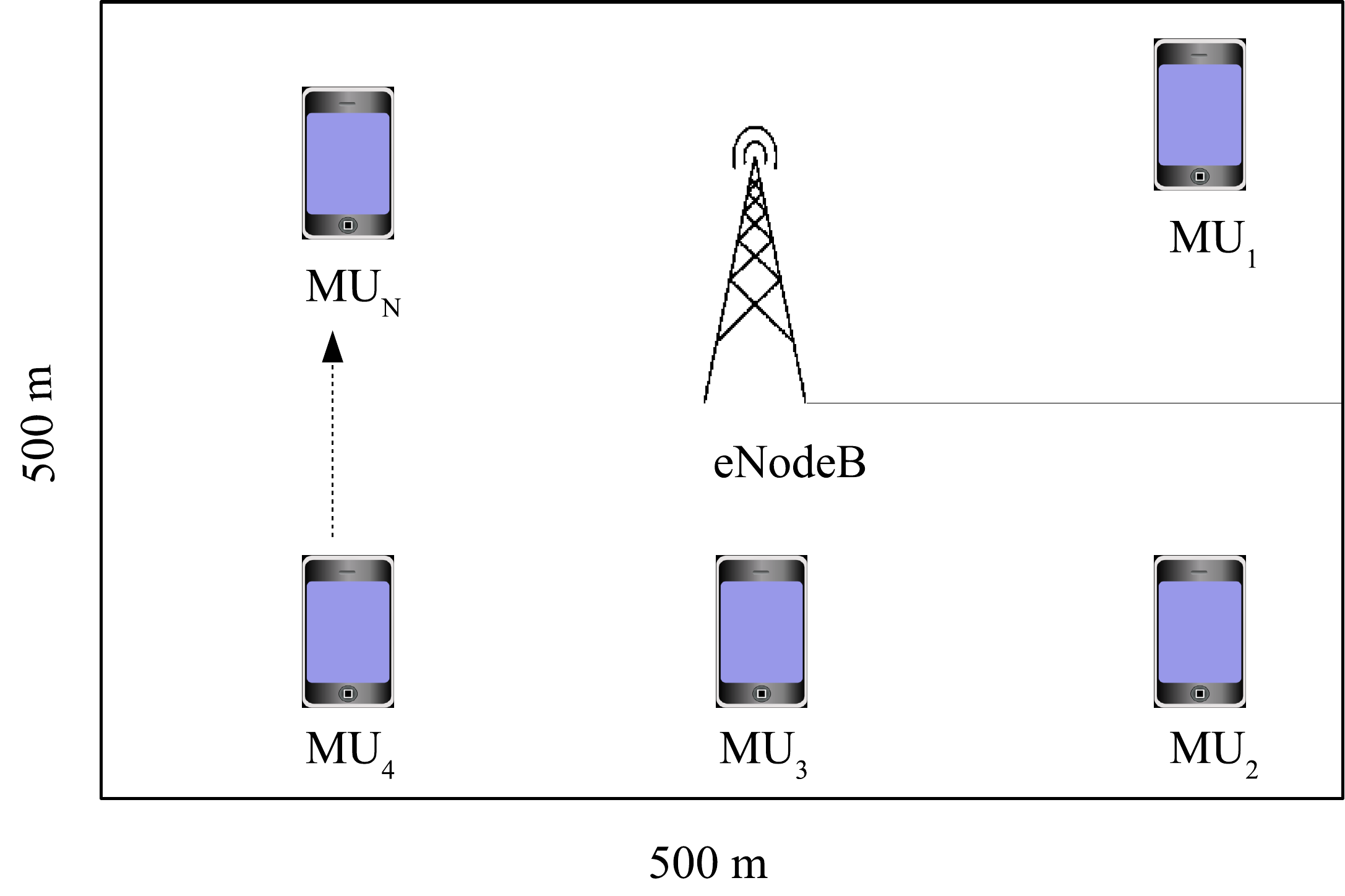}
	\caption{Network topology in non-congested scenario.}
	\label{fig:TopologyNormal}
\end{figure}

The energy saving algorithm consists of one eNodeB and the UEs that are uniformly distributed around the eNodeB as presented in Figure~\ref{fig:TopologyNormal}. All the simulation experiments setup have been adopted from LTE Release 8 \cite{ran2008requirements}. {These experiments have been conducted using the Vienna system-level simulator \cite{ikuno2010system},} in which the Multi-Input Multi-Output (MIMO) antennas and a system bandwidth of 5 MHz have been used. {The} simulation time in all experiments is 100 seconds, which is more than enough to show all stages of the system. As for updating the transmission power, it has been done at the beginning of each transmission time interval (TTI), where each second has 1000 TTIs, while the thermal noise density is set to -174 dBm/Hz during all simulation experiments. 

\begin{figure}[H]
	\centering
	\includegraphics[width=0.5\linewidth]{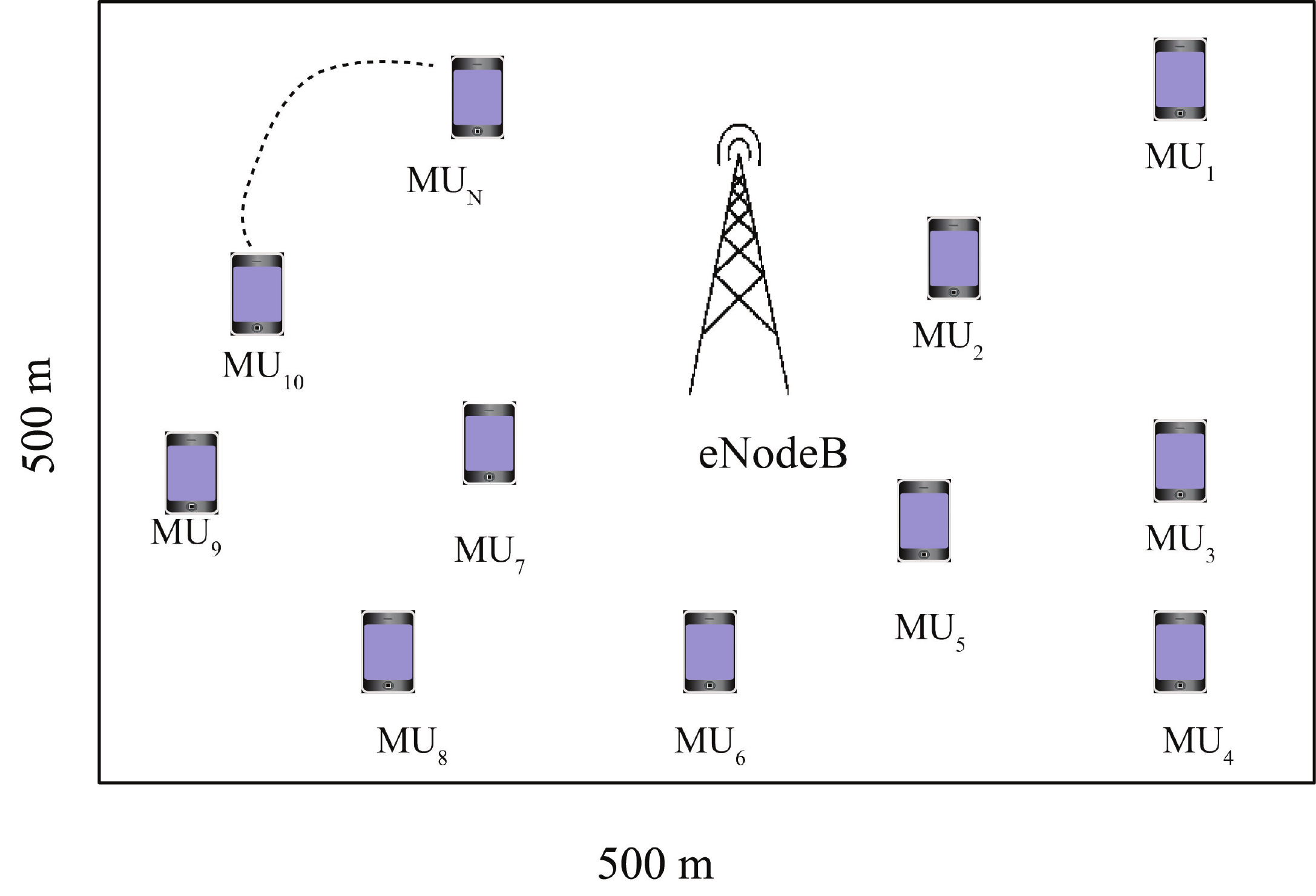}
	\caption{Network topology in congested scenario.}
	\label{fig:TopologyCongested}
\end{figure}

{The network topology used in this work has been adopted from \cite{haider2012energy}, which is used to evaluate the energy saving algorithm. As shown in Figure~\ref{fig:TopologyCongested},} the topology consists of one eNodeB with several Mobile Users (MUs) that are uniformly distributed in a congested network including MU$_{1}$, MU$_{2}$, ..., MU$_N$, where $N > 100$. Table \ref{table:1} shows the experimental setup of the simulation parameters.

\begin{specialtable}[H]
	 \tablesize{\small}
	\caption{Simulation parameters setup.}
	\setlength{\cellWidtha}{\columnwidth/2-2\tabcolsep+0.0in}
\setlength{\cellWidthb}{\columnwidth/2-2\tabcolsep+0.0in}
\scalebox{1}[1]{\begin{tabularx}{\columnwidth}{>{\PreserveBackslash\centering}m{\cellWidtha}>{\PreserveBackslash\centering}m{\cellWidthb}}
		\toprule
		\textbf{Parameter }					& \textbf{Value} 					\\ \midrule
		System bandwidth 			& 5 MHz 					\\ 
		Number of resource blocks	& 25 						\\ 
		Active sector of concern 	& eNodeB2-sector1 			\\ 
		TTI 		 				& 1 ms; 1000 TTI per second \\ 
		UE distribution 			& Uniform					\\ 
		Macroscopic path loss model & TS25814 					\\ 
		Simulation period 			& 100 {seconds}				\\ 
		Speed of the user 			& 4.16 m/s  	\\ 
		Transmission scheme 		& 2X2 MIMO, OLSM 			\\ 
		Cyclic prefix	 			& Normal 					\\ 
		Transmitter antenna gain 	& 18 dBi 					\\ 
		Receiver antenna gain 		& 0 dBi 					\\ 
		Maximum transmission power 	& 20 W 						\\ 
		Maximum delay 				& 20 ms 					\\ 
		Inter-eNodeB distance 		& 500 m 					\\ \bottomrule		
	\end{tabularx}	}
	\label{table:1}	
\end{specialtable}

\subsection{Impact of Varying Traffic Loads in Non-Congested Network}
\begin{figure}[H]
	\centering
	\includegraphics[width=0.55\linewidth, angle =-90]{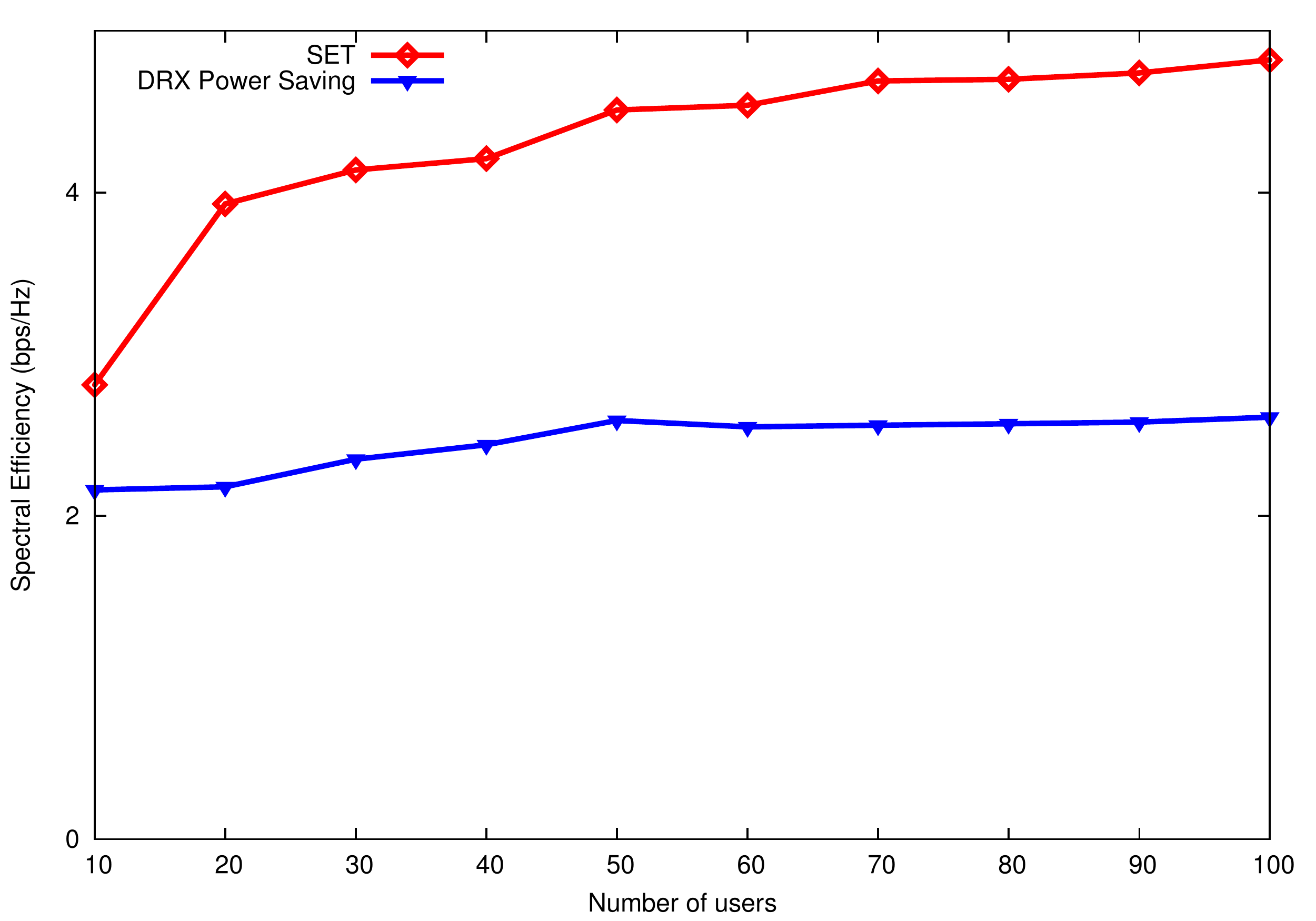}
	\caption{Spectral efficiency as the number of users increases.}
	\label{fig:NewSE}
\end{figure}

\begin{figure}[H]
	\centering
	\includegraphics[width=0.5\linewidth, angle =-90]{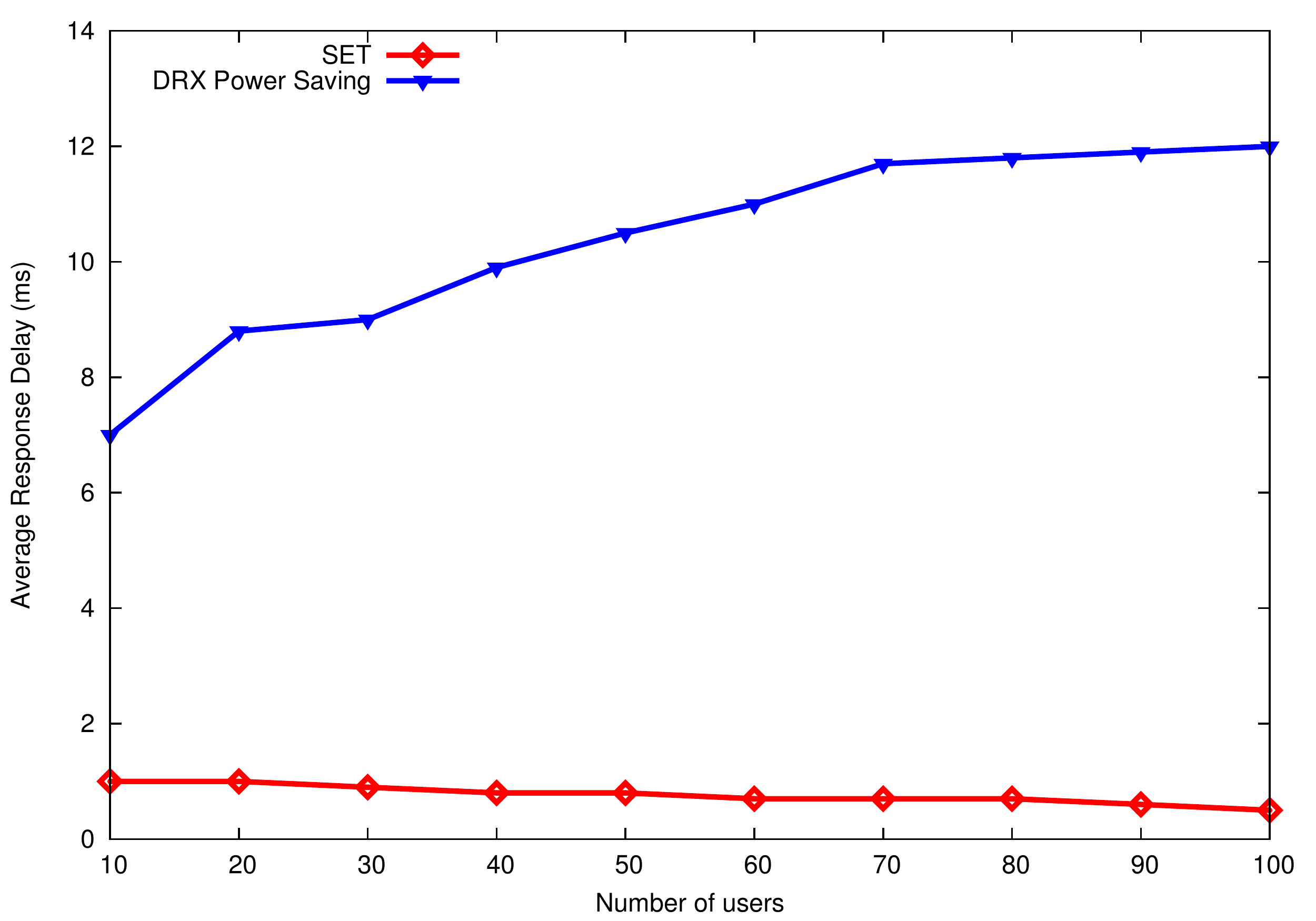}
	\caption{Average response delay as the number of users increases.}
	\label{fig:NewADelay}
\end{figure}

\begin{figure}[H]
	\centering
	\includegraphics[width=0.5\linewidth, angle =-90]{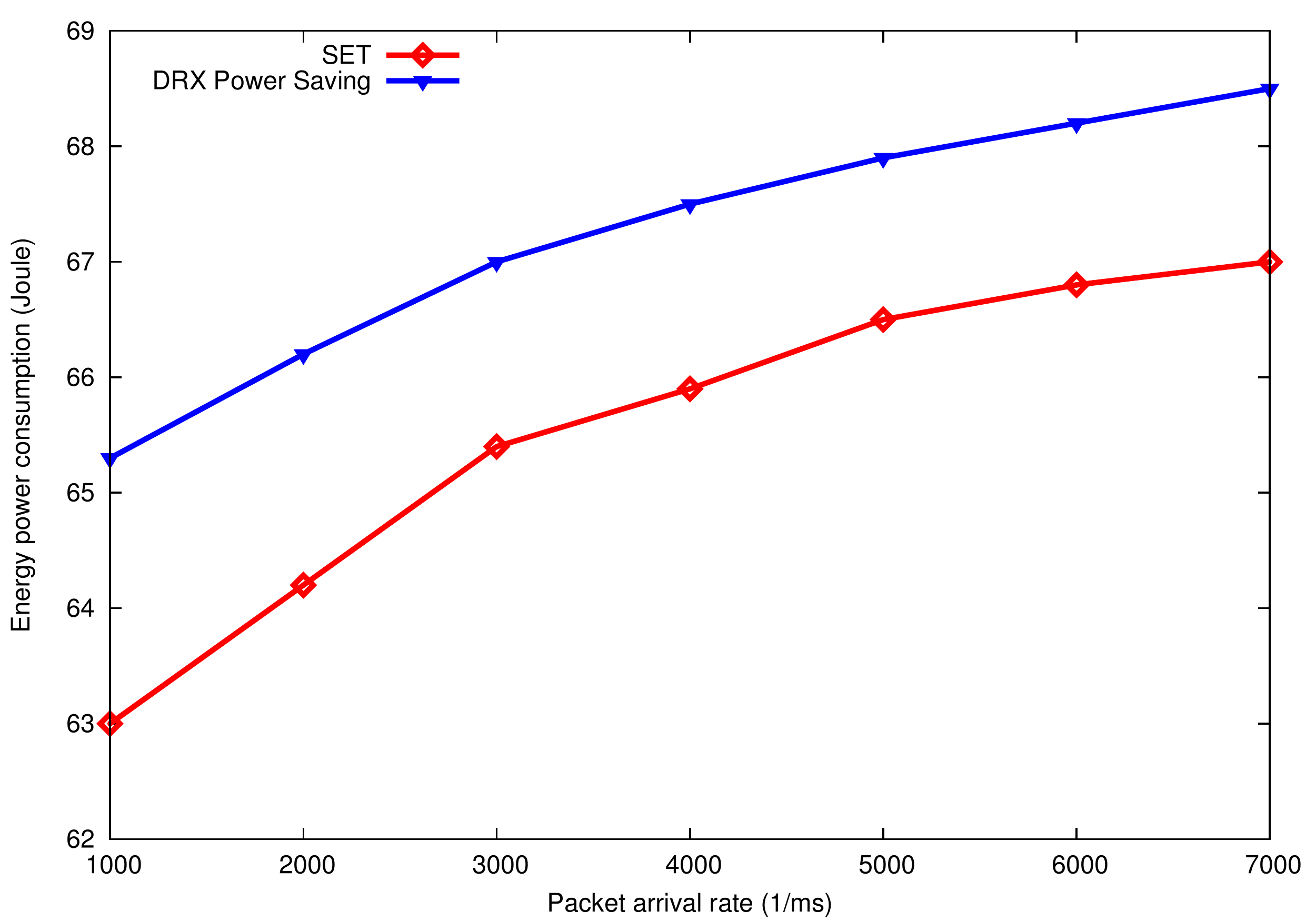}
	\caption{Energy consumption vs. packet arrival rate.}
	\label{fig:EneryCom}
\end{figure}

\begin{figure}[H]
	\centering
	\includegraphics[width=0.3\linewidth, angle =-90]{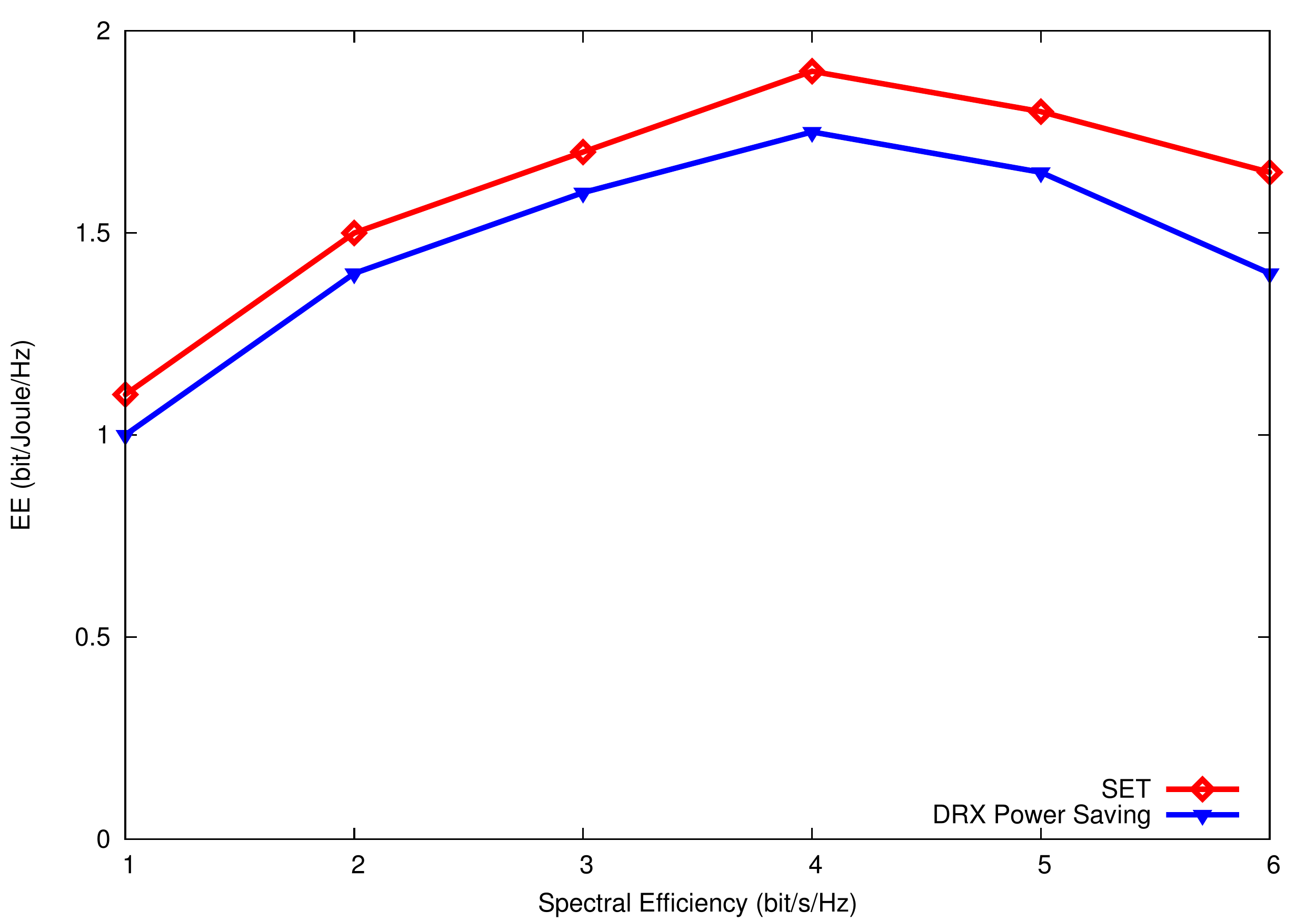}
	\caption{Effect of SE-EE 
		trade-off in non-congested network.}
	\label{fig:NormalSOP}
\end{figure}

In this experiment, a low density network with 100 pieces of UE has been used, where the UEs were uniformly distributed with an inter-eNodeB distance of 500 m. The objective here is to determine the network performance under an increasingly stressful scenario. {The results of the SE, average response delay, energy consumption, and the SE-EE trade-off are illustrated in Figures \ref{fig:NewSE}--\ref{fig:NormalSOP}, respectively}. 
{In Figure~\ref{fig:NewSE},} a comparison between the SET algorithm and the DRX power-saving algorithm has been presented in terms of the SE, by varying the number of users from 10 to 100. The proposed algorithm significantly outperforms the DRX algorithm in all cases because the larger share of resource blocks (RBs) is not lost due to users being under good channel condition before allocating resource blocks. As the number of users increases, the competition among users' resource blocks also increases. 

In addition, the use of transmission power to allocate resource blocks provides the proposed SET algorithm with a higher SE than that of the DRX power-saving algorithm. Furthermore, the DRX attempts to satisfy users while neglecting their conditions, which decreases its eNodeB performance. 

{The average response delay of the proposed SET algorithm compared to the DRX power-saving algorithm has been illustrated in Figure~\ref{fig:NewADelay}.} The average response delay of the DRX increases rapidly when the number of users increases because this algorithm does not consider the packet delay in resource block allocation. Furthermore, the proposed SET algorithm has lower average response delay than the DRX due to choosing the larger window before switching to the sleep-mode to accommodate incoming packets, which shows a significant improvement in the average response time in all cases.

{The effect on energy consumption by considering the packet arrival pattern in a non-congested network has been presented in Figure~\ref{fig:EneryCom}}. Compared to the DRX power-saving algorithm, the energy consumption of the proposed SET algorithm is lower due to the appropriate adjustment of eNodeB in the initial mode. Such an adjustment is performed by predicting the downlink packet arrival that is used to update the initial state. Consequently, minimizing the number of window sleep intervals reduces energy consumption. The proposed SET algorithm can prolong battery life by 81.92\% compared to only 18.08\% by the DRX algorithm.

Indeed, the proposed SET is utilized to obtain the Pareto optimal points between the SET algorithm and the DRX power-saving algorithm. In this manner, the MOP in Equation~\eqref{eq3} is transformed to the SOP in \mbox{Equation \eqref{eq5}}. As shown in Figure~\ref{fig:NormalSOP}, optimal EE increases until the saturation point with an increase in the SE. The proposed algorithm achieves considerable gain compared to the DRX algorithm, with optimal points of \mbox{1.9 bit/Hz} and 1.75 bit/Hz, respectively.

\subsection{Impact of Varying Traffic Loads in Congested Network}
In order to carry out this experiment, we implemented two scenarios: (1) varying network loads on a congested network using an omni-directional antenna. (2) varying network loads on a congested network using directional antennas. {Thus, the impact of using different types of antennas, such as omni and tilted directional antennas are presented in Sections \ref{omni} and \ref{directional}, respectively.}

\subsubsection{Impact of Varying Network Loads on a Congested Network Using an Omni-Directional Antenna} \label{omni}
\begin{figure}[H]
	\centering
	\includegraphics[width=0.5\linewidth, angle =-90]{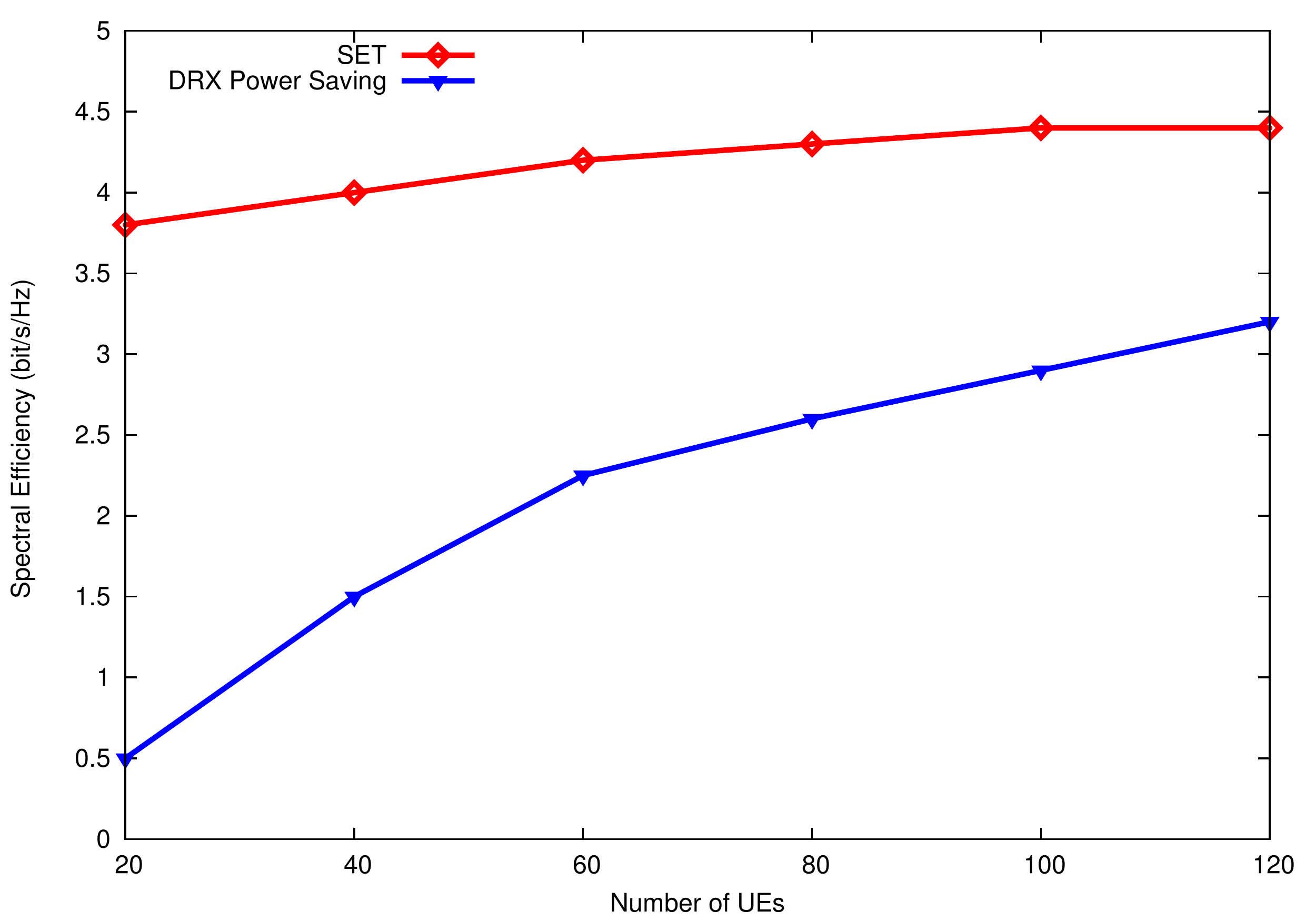}
	\caption{SE as the number of UEs are increased in congested network for with omni-directional antenna.}
	\label{fig:SpecOMi}
\end{figure}

\begin{figure}[H]
	\centering
	\includegraphics[width=0.3\linewidth, angle =-90]{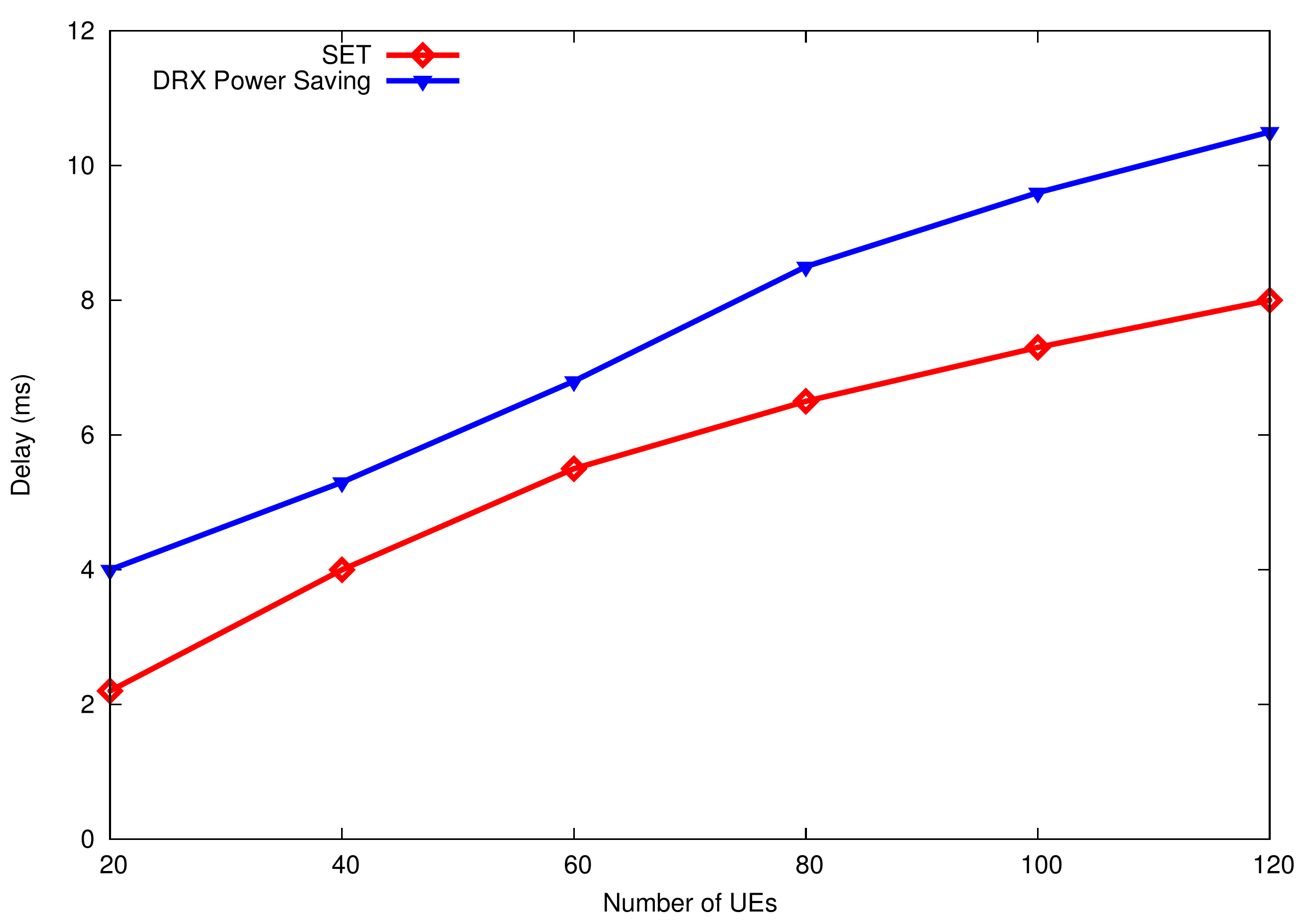}
	\caption{Delay as the number of UEs 
		are increased in congested network for with omni-directional antenna.}
	\label{fig:DelayOmni}
\end{figure}

\begin{figure}[H]
	\centering
	\includegraphics[width=0.5\linewidth, angle =-90]{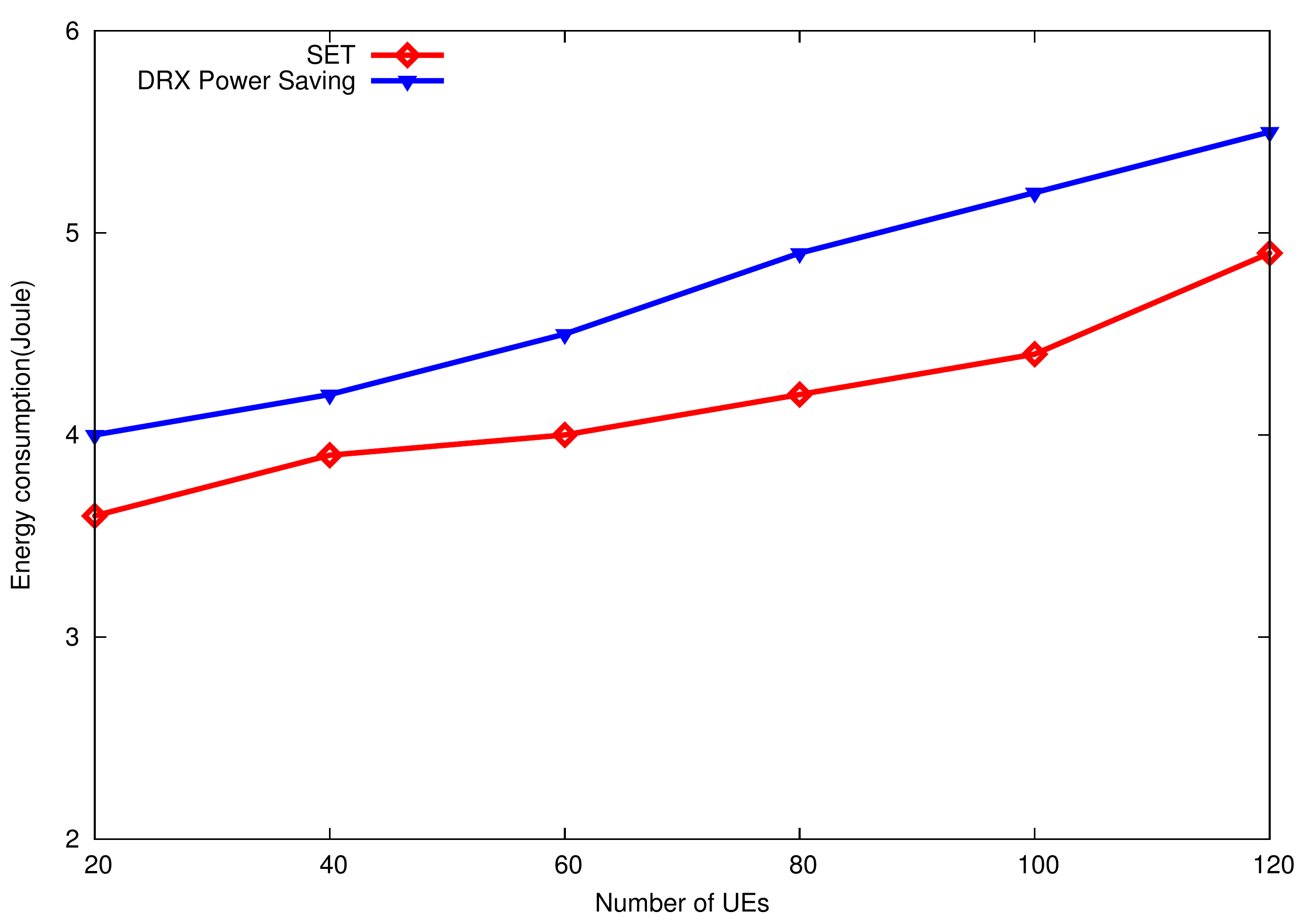}
	\caption{Energy consumption by increasing number of UEs in congested network for with omni-directional antenna.}
	\label{fig:EnergyOmni}
\end{figure}

\begin{figure}[H]
	\centering
	\includegraphics[width=0.5\linewidth, angle =-90]{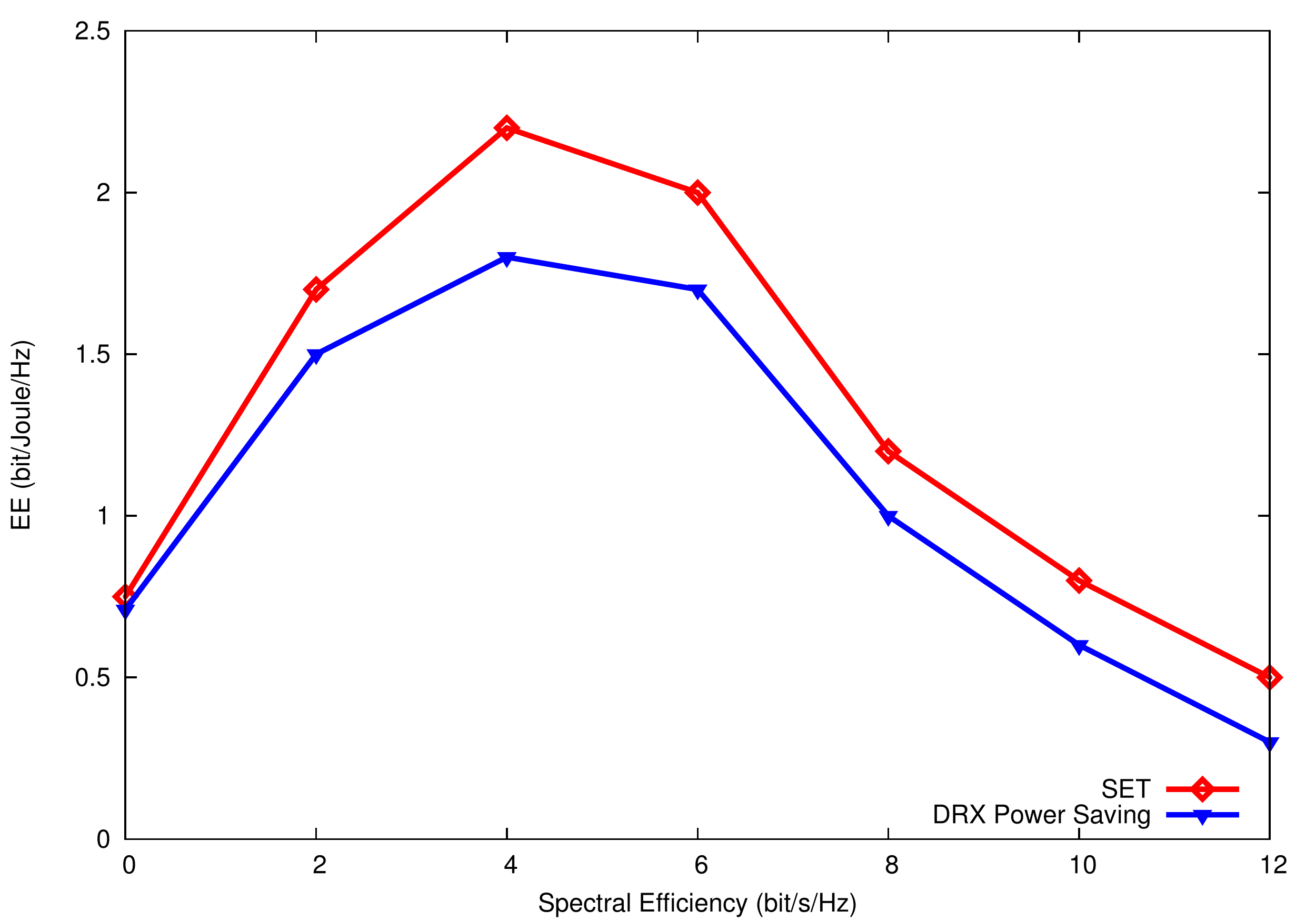}
	\caption{SE-EE trade-off of congested network with omni-directional antenna.}
	\label{fig:OmniSOP}
\end{figure}

An omni-directional antenna that spreads radio signals in all directions has been used in a congested network scenario that contains a large number of UE $>$ 100 \cite{wang2016service}. The pieces of UE are uniformly distributed with an inter-eNodeB distance of 500 m. The objective of implementing this scenario is to determine the network performance under an increasingly stressful condition, where Figures \ref{fig:SpecOMi}--\ref{fig:OmniSOP} present the results of the SE, delay, energy consumption, and SET, respectively.

{A comparison of the SE between the proposed SET algorithm and the DRX power-saving algorithm has been shown in Figure~\ref{fig:SpecOMi}}. As explained in Section \ref{SET}, the proposed SET algorithm allocates the RBs to users with good channel states, which exhibits better performance than the DRX algorithm. It worth noting that disregarding the selection of an appropriate channel state by the DRX algorithm results in poor SE, which affects the performance of the eNodeB.

{A comparison between the delay of the proposed SET algorithm to that of the DRX power-saving algorithm has been presented in Figure~\ref{fig:DelayOmni}}. The DRX algorithm exhibits an increased delay because the RBs allocation is established while disregarding the packet delay requirements. Moreover, the proposed algorithm is able to minimize the delay because it uses an appropriate selection of window size. Given the frequent arrival of users and the spread of radio signals in different directions by the omni-directional antenna, the competition among users to access the radio signals increases the delay in the case of the DRX algorithm that, consequently, increases the interference level.

Subsequently, the energy consumption of the eNodeB is investigated, where \mbox{Figure~\ref{fig:EnergyOmni}} shows that increasing the number of UE in the network environment will cause a rise in the interference, thereby increasing the energy consumption, in which the increase of interference makes it difficult to find an optimal power transmission. For this reason, the proposed algorithm outperforms the DRX power-saving algorithm with the least energy consumption due to its ability to adjust eNodeB by predicting the arrival pattern of user requests.

{The impact of increasing the number of UE pieces in a congested network using omni-directional antenna by utilizing SET has been presented in Figure~\ref{fig:OmniSOP},} in which the increase in the SE will always increase the EE.

The proposed SET algorithm achieves the optimal point at 2.2 bit/s/Hz, which is higher than that of the DRX power-saving algorithm whose optimal point was at \mbox{1.78 bit/s/Hz}. This result indicates that the proposed SET algorithm outperforms the DRX algorithm because it can achieve optimal value with reduced energy consumption.

\subsubsection{Impact of Varying Network Loads on a Congested Network Using a Directional Antenna}  \label{directional}
A directional antenna that spreads radio signal in a specific direction called 'sector', i.e., either $60^\circ$ or $120^\circ$ has been used in this scenario. In fact, transmitting or receiving radio signals in one direction or specific angle leads to a low level of interference, which consequently leads to improving the network capacity and coverage that in turn increase its successful delivery rate between UE and eNodeB/antenna.

\begin{figure}[H]
	\centering
	\includegraphics[width=0.5\linewidth, angle =-90]{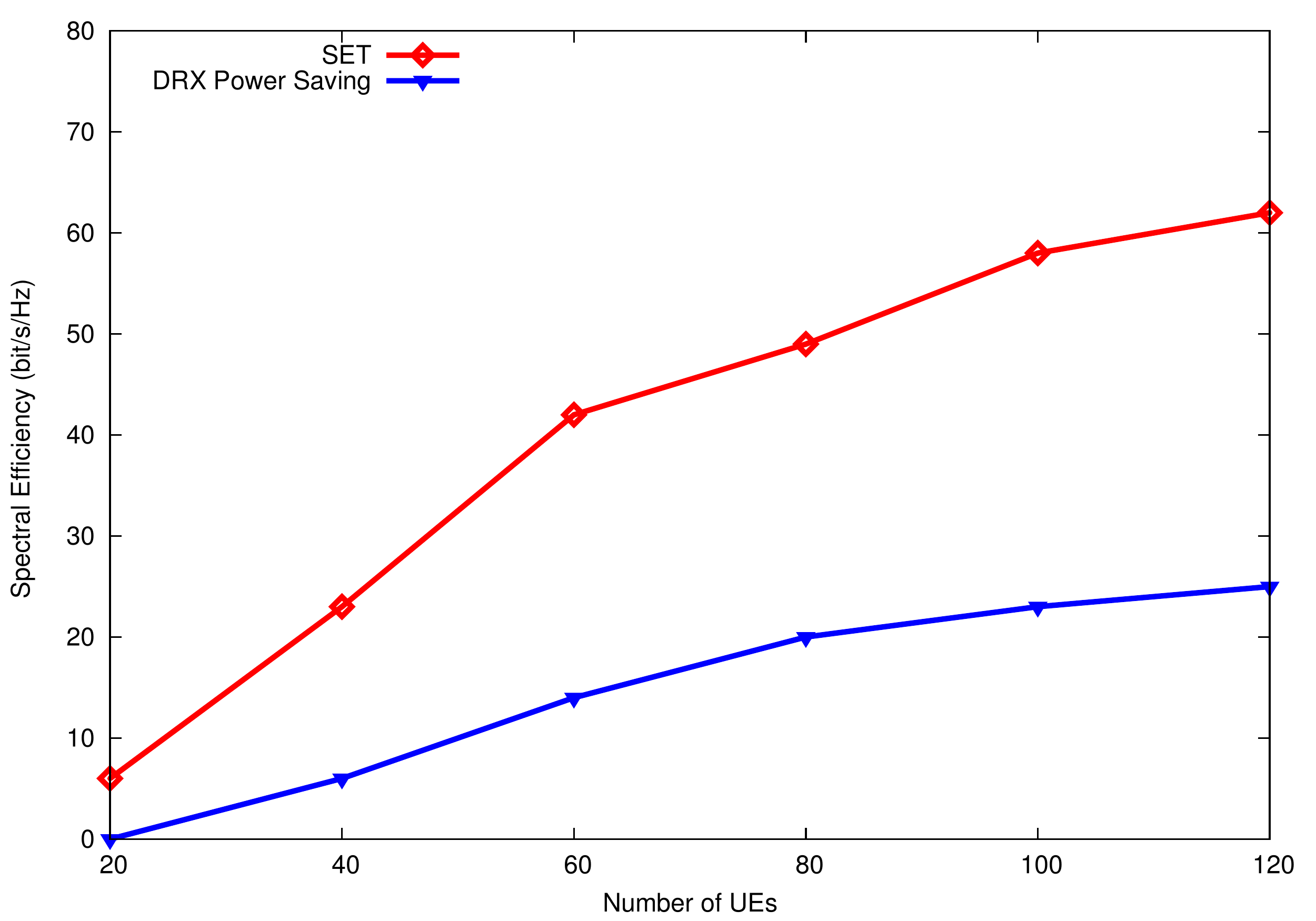}
	\caption{SE as the number of UEs are increased in congested network for with directional antenna.}
	\label{fig:SpectODir}
\end{figure}

Moreover, Figure~\ref{fig:SpectODir} shows how manipulating the number of UEs per eNodeB could improve the SE, where increasing the number of UEs increases the performance due to the competition. Since cell sectoring is performed by the proposed algorithm to enable radio signals to converge at a certain degree, thus, cell edge users are served with RBs because their channel state is in good condition. For this reason, the proposed algorithm exhibits a higher gain than the DRX power-saving algorithm due to its ability to rotate a directional antenna, which enables it to cope with the demands of cell edge users.

\begin{figure}[H]
	\centering
	\includegraphics[width=0.5\linewidth, angle =-90]{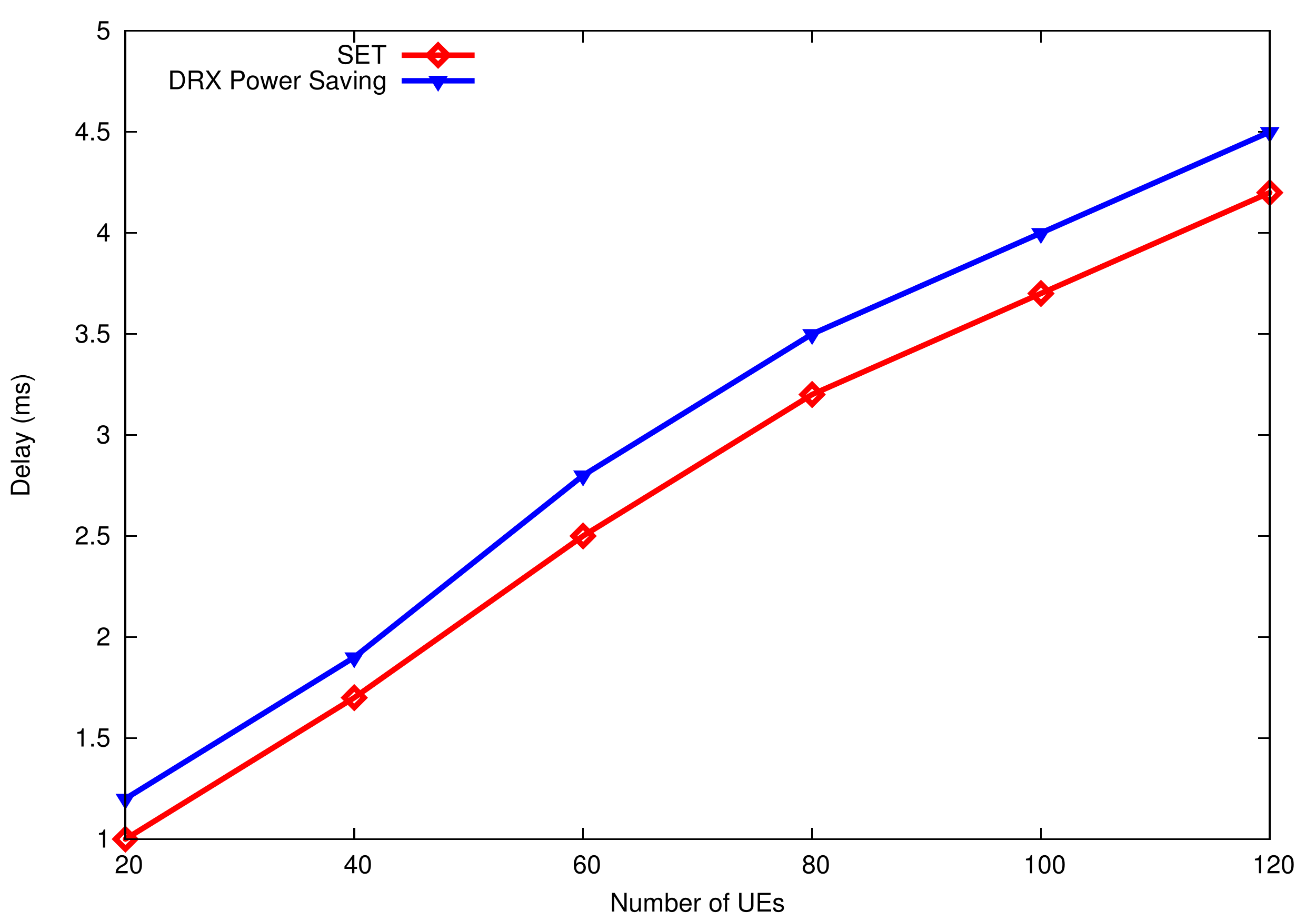}
	\caption{Delay as the number of UEs are increased in congested network for with directional antenna.}
	\label{fig:DelyDirect}
\end{figure}

In Figure~\ref{fig:DelyDirect}, a comparison between the delay of the proposed SET algorithm and that of the DRX power-saving algorithm has been presented, where the latter has increased delay due to disregarding the packet delay requirements during RB allocation. The proposed algorithm was able to reduce the delay due to selecting the appropriate window size. Moreover, the sectoring approach of a directional antenna considerably affects the delay metric due to the use of a reduced coverage area, however, the proposed algorithm was able to minimize the delay, which minimizes the interference. Meanwhile, the DRX algorithm increases the delay that significantly led to high energy consumption.

\begin{figure}[H]
	\centering
	\includegraphics[width=0.5\linewidth, angle =-90]{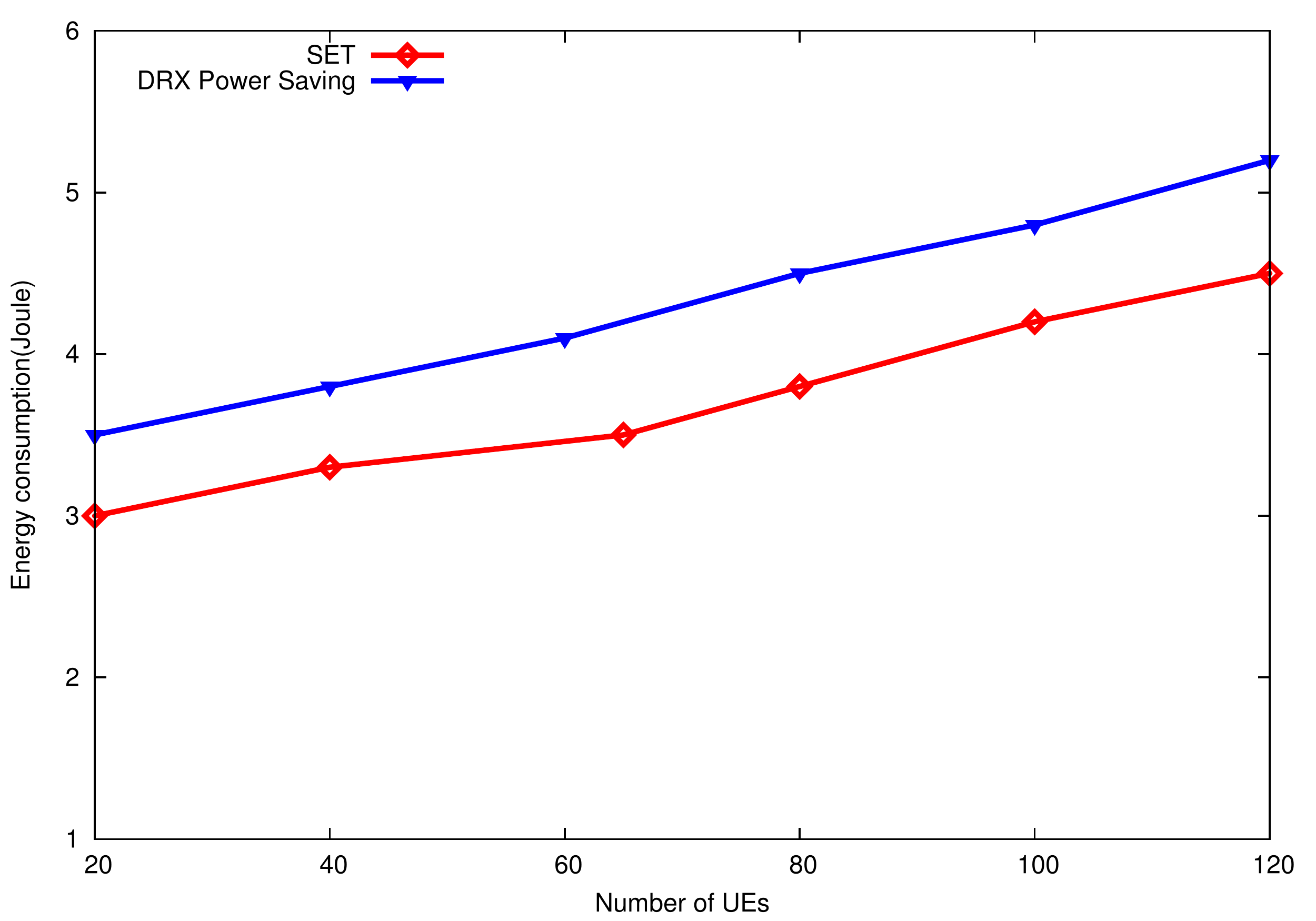}
	\caption{Energy consumption as the number of UEs are increased in congested network for with directional antenna.}
	\label{fig:EnergyDirectional}
\end{figure}

{The increase in the number of UE increases the amount of consumed energy due to the high interference, as shown in Figure~\ref{fig:EnergyDirectional}}. Consequently, the total performance is degraded when the energy consumption at eNodeB increases, which leads to a significant delay increase with both algorithms. However, the proposed algorithm still able to outperform the DRX power-saving algorithm due to its appropriate method to adjust the eNodeB to a certain degree.

\begin{figure}[H]
	\centering
	\includegraphics[width=0.3\linewidth, angle =-90]{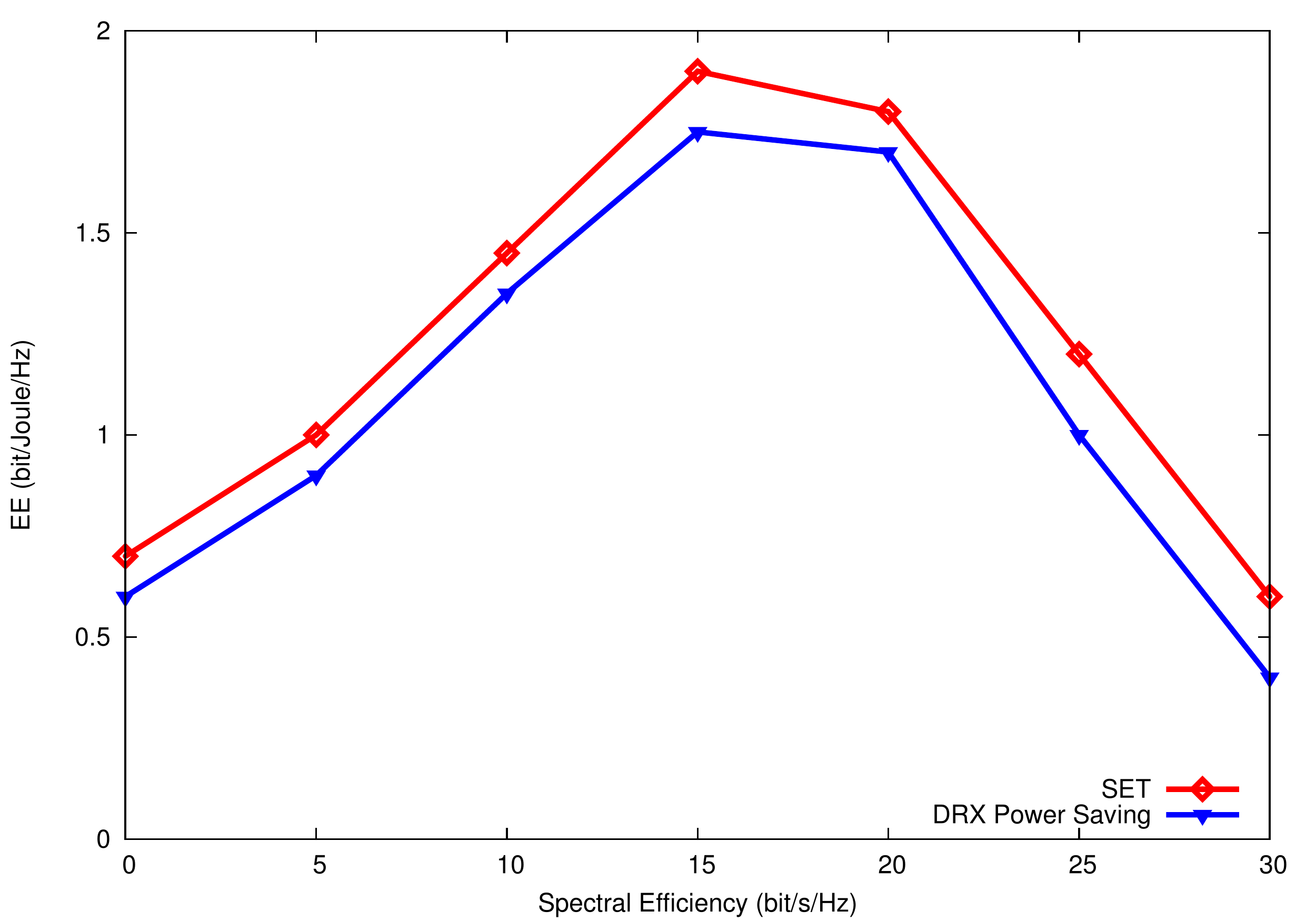}
	\caption{SE-EE trade-off of congested network with directional antenna.}
	\label{fig:DRSOP}
\end{figure}

{The impact of increasing the number of UEs in a congested network using a directional antenna by utilizing SET has been shown in Figure~\ref{fig:DRSOP}}. This figure shows that the increase in SE always leads to EE increase. The proposed algorithm achieves its optimal point at \mbox{1.9 bit/s/Hz}, which is higher than that of the DRX power-saving algorithm at \mbox{1.75 bit/s/Hz}. This finding indicates that the proposed algorithm is able to outperform the DRX algorithm due to achieving its optimal value with lower energy consumption.

\section{Conclusions}\label{Con}
This research addressed the high energy consumption problem in modern mobile communication systems by controlling the transmission power of eNodeB. The two parameters that play a vital role in improving energy efficiency are the adaptive initial and final thresholds, where these two parameters were adjusted by considering a stochastic traffic arrival pattern of UEs. In addition, the proposed SET algorithm was formulated as a MOP to determine the trade-off between the SE and the EE. Similarly, an interference approach was developed by applying antenna patterns to provide efficient energy management at eNodeB. Several simulation experiments were conducted, which demonstrated the ability of the proposed algorithm to outperform the DRX power-saving algorithm in terms of SE, average response delay, and energy consumption.

\authorcontributions{All authors of this article have contributed to the work as follows: conceptualization, validation, formal analysis, methodology, software, visualization, and original draft preparation have been done by M.M.; technical review, proofreading, editing, and writing the final draft have been done by M.A.A.; project administration, supervision, and funding acquisition have been done by Z.M.H. All authors have read and agreed to the published version of the manuscript.}

\funding{This work is funded by Universiti Putra Malaysia under Putra Berimpak Grant number (9659400).}

\institutionalreview{Not applicable, because the study does not involving humans or animals.}

\informedconsent{Not applicable, because this study not involving humans.}

\dataavailability{Not applicable}

\acknowledgments{\textls[-30]{This work is supported by Universiti Putra Malaysia.}}

\conflictsofinterest{The authors declare no conflict of interest.} 

\end{paracol}
\reftitle{References}

\begin{thebibliography}{33}
	\providecommand{\natexlab}[1]{#1}
	
	\bibitem[Atayero \em{et~al.}(2011)Atayero, Luka, Orya, and
	Iruemi]{atayero20113gpp}
	Atayero, A.A.; Luka, M.K.; Orya, M.K.; Iruemi, J.O.
	\newblock 3GPP long term evolution: Architecture, protocols and interfaces.
	\newblock {\em Int. J. Inf. Commun. Res.} {\bf 2011}, {\em 1},~306--310.
	
	\bibitem[Holma and Toskala(2012)]{holma2012lte}
	Holma, H.; Toskala, A.
	\newblock {\em LTE Advanced: 3GPP Solution for IMT-Advanced}; John Wiley \&
	Sons: 2012.
	
	\bibitem[Tung \em{et~al.}(2015)Tung, Wang, Hsueh, and Chang]{tung2015analysis}
	Tung, L.P.; Wang, L.C.; Hsueh, C.W.; Chang, C.J.
	\newblock Analysis of DRX power saving with RRC states transition in LTE
	networks.
	\newblock   In Proceedings of the 2015 European Conference on Networks and Communications (EuCNC), \mbox{pp. 301--305.} IEEE, 2015.
	
	\bibitem[Tang \em{et~al.}(2014)Tang, So, Alsusa, and Hamdi]{tang2014resource}
	Tang, J.; So, D.K.; Alsusa, E.; Hamdi, K.A.
	\newblock Resource efficiency: A new paradigm on energy efficiency and spectral
	efficiency tradeoff.
	\newblock {\em IEEE Trans. Wirel. Commun.} {\bf 2014}, {\em
		13},~4656--4669.
	
	\bibitem[Chen \em{et~al.}(2015)Chen, Hu, Wu, and Li]{chen2015tradeoff}
	Chen, X.; Hu, R.Q.; Wu, G.; Li, Q.C.
	\newblock Tradeoff between energy efficiency and spectral efficiency in a delay
	constrained wireless system.
	\newblock {\em Wirel. Commun. Mob. Comput.} {\bf 2015}, {\em
		15},~1945--1956.
	
	\bibitem[Souza \em{et~al.}(2015)Souza, Abr{\~a}o, Sampaio, Jeszensky,
	P{\'e}rez-Romero, and Casadevall]{souza2015energy}
	Souza, {\'A}.R.; Abr{\~a}o, T.; Sampaio, L.H.; Jeszensky, P.J.E.;
	P{\'e}rez-Romero, J.; Casadevall, F.
	\newblock Energy and spectral efficiencies trade-off with filter optimisation
	in multiple access interference-aware networks.
	\newblock {\em Trans. Emerg. Telecommun. Technol.} {\bf
		2015}, {\em 26},~670--685.
	
	\bibitem[Abr{\~a}o \em{et~al.}(2016)Abr{\~a}o, Sampaio, Yang, Cheung,
	Jeszensky, and Hanzo]{abrao2016energy}
	Abr{\~a}o, T.; Sampaio, L.D.H.; Yang, S.; Cheung, K.T.K.; Jeszensky, P.J.E.;
	Hanzo, L.
	\newblock Energy efficient OFDMA networks maintaining statistical QoS
	guarantees for delay-sensitive traffic.
	\newblock {\em IEEE Access} {\bf 2016}, {\em 4},~774--791.
	
	\bibitem[Son \em{et~al.}(2015)Son, Oh, and Krishnamachari]{son2015energy}
	Son, K.; Oh, E.; Krishnamachari, B.
	\newblock Energy-efficient design of heterogeneous cellular networks from
	deployment to operation.
	\newblock {\em Comput. Netw.} {\bf 2015}, {\em 78},~95--106.
	
	\bibitem[Song \em{et~al.}(2015)Song, Ni, Navaie, Hou, and Wu]{song2015energy}
	Song, Z.; Ni, Q.; Navaie, K.; Hou, S.; Wu, S.
	\newblock Energy-and spectral-efficiency tradeoff with alpha-fairness in
	downlink OFDMA systems.
	\newblock {\em IEEE Commun. Lett.} {\bf 2015}, {\em 19},~1265--1268.
	
	\bibitem[Pervaiz \em{et~al.}(2015)Pervaiz, Musavian, Ni, and
	Ding]{pervaiz2015energy}
	Pervaiz, H.; Musavian, L.; Ni, Q.; Ding, Z.
	\newblock Energy and spectrum efficient transmission techniques under QoS
	constraints toward green heterogeneous networks.
	\newblock {\em IEEE Access} {\bf 2015}, {\em 3},~1655--1671.
	
	\bibitem[Li \em{et~al.}(2015)Li, Jiang, Pan, Liu, and You]{li2015energy}
	Li, Z.; Jiang, H.; Pan, Z.; Liu, N.; You, X.
	\newblock Energy spectral efficiency tradeoff in downlink OFDMA network.
	\newblock {\em Int. J. Commun. Syst.} {\bf 2015}, {\em
		28},~1450--1461.
	
	\bibitem[Kim \em{et~al.}(2008)Kim, Kang, and Choi]{kim2008remaining}
	Kim, M.G.; Kang, M.; Choi, J.
	\newblock Remaining energy-aware power management mechanism in the 802.16 e
	MAC.
	\newblock  In Proceedings of the Consumer Communications and Networking Conference, Las Vegas, NV, USA, 10--12 Jannuary 2008; pp. 222--226.
	
	\bibitem[Richter \em{et~al.}(2009)Richter, Fehske, and
	Fettweis]{richter2009energy}
	Richter, F.; Fehske, A.J.; Fettweis, G.P.
	\newblock Energy efficiency aspects of base station deployment strategies for 
	cellular networks.
	\newblock  In Proceedings of the Vehicular Technology Conference Fall (VTC 2009-Fall), pp. 1--5, IEEE, 2009.
	
	\bibitem[Ling \em{et~al.}(2010)Ling, Wang, Wang, and Shi]{ling2010schemes}
	Ling, L.; Wang, T.; Wang, Y.; Shi, C.
	\newblock Schemes of power allocation and antenna port selection in OFDM
	distributed antenna systems.
	\newblock  In Proceedings of the Vehicular Technology Conference Fall (VTC 2010-Fall), pp. 1--5, IEEE, 2010.
	
	\bibitem[Isheden and Fettweis(2011)]{isheden2011energy}
	Isheden, C.; Fettweis, G.P.
	\newblock Energy-efficient link adaptation with transmitter CSI.
	\newblock  In Proceedings of the Wireless Communications and Networking Conference (WCNC), pp. 1381--1386, IEEE, 2011.
	
	\bibitem[Haider \em{et~al.}(2012)Haider, Wang, Haas, Hepsaydir, and
	Ge]{haider2012energy}
	Haider, F.; Wang, C.X.; Haas, H.; Hepsaydir, E.; Ge, X.
	\newblock Energy-efficient subcarrier-and-bit allocation in multi-user OFDMA
	systems.
	\newblock  In Proceedings of the Vehicular Technology Conference (VTC Spring), pp. 1--5, IEEE, 2012.
	
	\bibitem[Khakurel \em{et~al.}(2013)Khakurel, Musavian, and
	Le-Ngoc]{khakurel2013trade}
	Khakurel, S.; Musavian, L.; Le-Ngoc, T.
	\newblock Trade-off between spectral and energy efficiencies in a fading
	communication link.
	\newblock  In Proceedings of the Vehicular Technology Conference (VTC Spring), pp. 1--5, IEEE, 2013.
	
	\bibitem[Coskun and Ayanoglu(2017)]{coskun2017energy}
	Coskun, C.C.; Ayanoglu, E.
	\newblock Energy-spectral efficiency tradeoff for heterogeneous networks with
	QoS constraints.
	\newblock  In Proceedings of the Communications (ICC), 
	pp. 1--7, IEEE, 2017.
	
	\bibitem[Tsilimantos \em{et~al.}(2016)Tsilimantos, Gorce, Jaffr{\`e}s-Runser,
	and Poor]{tsilimantos2016spectral}
	Tsilimantos, D.; Gorce, J.M.; Jaffr{\`e}s-Runser, K.; Poor, H.V.
	\newblock Spectral and energy efficiency trade-offs in cellular networks.
	\newblock {\em IEEE Trans. Wirel. Commun.} {\bf 2016}, {\em
		15},~54--66.
	
	\bibitem[Li \em{et~al.}(2017)Li, Jiang, Li, Pan, Liu, and You]{li2017energy}
	Li, Z.; Jiang, H.; Li, P.; Pan, Z.; Liu, N.; You, X.
	\newblock Energy-Spectral-Efficiency Tradeoff in Interference-Limited Wireless
	Networks.
	\newblock {\em Wirel. Pers. Commun.} {\bf 2017}, {\em
		96},~5515--5532.
	
	\bibitem[Salman \em{et~al.}(2018)Salman, Mansoor, Jalab, Sabri, and
	Ahmed]{salman2018joint}
	Salman, M.I.; Mansoor, A.M.; Jalab, H.A.; Sabri, A.Q.M.; Ahmed, R.
	\newblock A Joint Evaluation of Energy-Efficient Downlink Scheduling and
	Partial CQI Feedback for LTE Video Transmission.
	\newblock {\em Wirel. Pers. Commun.} {\bf 2018}, {\em
		98},~189--211.
	
	\bibitem[ETSI(2010)]{LTE2010specification}
	ETSI.
	\newblock 3GPP TS 36.814 Evolved Universal Terrestrial Radio Access (E-UTRA);
	Further advancements for E-UTRA physical layer aspects (Release 9), 2010.
	
	\bibitem[Emmerich and Deutz(2006)]{emmerich2006multicriteria}
	Emmerich, M.; Deutz, A.
	\newblock {\em Multicriteria Optimization and Decision Making}, principles,
	algorithms and case studies ed.; LIACS; Leiden University, 2006.
	
	\bibitem[Coello \em{et~al.}(2007)Coello, Lamont, Van~Veldhuizen,
	et~al.]{coello2007evolutionary}
	Coello, C.A.C.; Lamont, G.B.; Van~Veldhuizen, D.A.
	\newblock {\em Evolutionary Algorithms for Solving Multi-Objective Problems};
	Volume~5, pp.~79--104, New York: Springer, 2007.
	
	\bibitem[Marler and Arora(2004)]{marler2004survey}
	Marler, R.T.; Arora, J.S.
	\newblock Survey of multi-objective optimization methods for engineering.
	\newblock {\em Struct. Multidiscip. Optim.} {\bf 2004}, {\em
		26},~369--395.
	
	\bibitem[Lee and Sohn(2017)]{lee2017distributed}
	Lee, S.H.; Sohn, I.
	\newblock Distributed energy-saving cellular network management using
	message-passing.
	\newblock {\em IEEE Trans. Veh. Technol.} {\bf 2017}, {\em
		66},~635--644.
	
	\bibitem[Liu \em{et~al.}(2018)Liu, Xiong, Yu, Feng, Li, Qiu, and
	Wang]{liu2018energy}
	Liu, L.; Xiong, A.; Yu, P.; Feng, L.; Li, W.; Qiu, X.; Wang, M.
	\newblock Energy-saving management mechanism based on hybrid energy supplies in
	multi-operator shared LTE networks.
	\newblock  In Proceedings of the NOMS 2018-2018 IEEE/IFIP Network Operations and Management
	Symposium, pp.~1--5. IEEE, 2018.
	
	\bibitem[Auer \em{et~al.}(2011)Auer, Giannini, Desset, Godor, Skillermark,
	Olsson, Imran, Sabella, Gonzalez, Blume, et~al.]{auer2011much}
	Auer, G.; Giannini, V.; Desset, C.; Godor, I.; Skillermark, P.; Olsson, M.;
	Imran, M.A.; Sabella, D.; Gonzalez, M.J.; Blume, O.; et~al.
	\newblock How much energy is needed to run a wireless network?
	\newblock {\em IEEE Wirel. Commun.} {\bf 2011}, {\em 18(5)}, pp.~40--49.
	
	\bibitem[Jensen \em{et~al.}(2012)Jensen, Lauridsen, Mogensen, S{\o}rensen, and
	Jensen]{jensen2012lte}
	Jensen, A.R.; Lauridsen, M.; Mogensen, P.; S{\o}rensen, T.B.; Jensen, P.
	\newblock LTE UE power consumption model.
	\newblock  In Proceedings of the 2012 IEEE Vehicular Technology Conference (VTC Fall), pp,~1--5, IEEE, 2012.
	
	
	
	\bibitem[Wang and Hsieh(2016)]{wang2016service}
	Wang, Y.C.; Hsieh, S.Y.
	\newblock Service-differentiated downlink flow scheduling to support QoS in
	long term evolution.
	\newblock {\em Comput. Netw.} {\bf 2016}, {\em 94},~344--359.
	
	\bibitem[Elwekeil \em{et~al.}(2017)Elwekeil, Alghoniemy, Muta, Abdel-Rahman,
	Gacanin, and Furukawa]{elwekeil2017performance}
	Elwekeil, M.; Alghoniemy, M.; Muta, O.; Abdel-Rahman, A.B.; Gacanin, H.;
	Furukawa, H.
	\newblock Performance evaluation of an adaptive self-organizing frequency reuse
	approach for OFDMA downlink.
	\newblock {\em Wirel. Netw.} {\bf 2017}, \emph{25}, 507--519.
	
	\bibitem[RAN(2008)]{ran2008requirements}
	TAJIKA, Y.; TAOKA, H. and HIGUCHI, K.
	\newblock 3GPP; TSG RAN; Requirements for further advancements for E-UTRA (LTE-Advanced) 3GPP; TSG RAN; Requirements for further advancements for E-UTRA (LTE-Advanced).
	\newblock IEICE transactions on communications, \emph{94(12)}, pp.~3280--3288, 2008.

	
	\bibitem[Ikuno \em{et~al.}(2010)Ikuno, Wrulich, and Rupp]{ikuno2010system}
	Ikuno, J.C.; Wrulich, M.; Rupp, M.
	\newblock System level simulation of LTE networks.
	\newblock In Proceedings of the 2010 IEEE 71st Vehicular Technology Conference (VTC 2010-Spring), pp.~1--5, 2010.
	
\end{thebibliography}


\end{document}